\documentstyle[prb,aps,12pt]{revtex}
\begin{document}
\title{Infrared absorption of the charge-ordering phase: Lattice effects}
\author{C.A. Perroni, V. Cataudella, G. De Filippis, G. Iadonisi, V. 
Marigliano Ramaglia, and F. Ventriglia}
\address{Coherentia-INFM  and Dipartimento di Scienze Fisiche, \\
Universit\`{a} degli Studi di Napoli ``Federico II'',\\
Complesso Universitario Monte Sant'Angelo,\\
Via Cintia, I-80126 Napoli, Italy}
\date{\today}
\maketitle

\begin{abstract}
The phase diagram of the half-filled spinless Holstein model for electrons 
interacting with quantum phonons is derived in three dimensions  
extending at finite temperature $T$ a variational approach introduced for the 
one-dimensional $T=0$ case. 
Employing the variational scheme, the spectral and optical properties
of the system are evaluated in the different regimes that characterize
the normal and ordered state.
The effects of the charge-ordering ($CO$) induce a transfer of spectral weight
from low to high energies in the conductivity spectra, as the temperature 
decreases or the strength of the electron-phonon ($el-ph$) interaction 
increases.    
The inclusion of effects of lattice fluctuations is able to smooth the inverse 
square-root singularity expected for the case of the mean-field approach and 
determines a subgap tail absorption.
Moreover, in the weak to intermediate coupling regime, a two-component
structure is obtained within the $CO$ phase at low frequency: 
the remnant Drude-like term and the incipient absorption band centered 
around the gap energy.
\end{abstract}


\newpage 
In the last years there has been a renewed interest in charge density wave 
($CDW$) materials. \cite{gruner} The transition to a $CO$ phase is common to a
 wide range of compounds, \cite{springer}  including quasi-one-dimensional 
organic conductors, \cite{gruner,kago} dichalcogenides, \cite{wilson,mcwhan} 
molybdenum bronzes \cite{springer,travi} and A-15 materials. \cite{testardi} 
Moreover, the $CO$ transition has been associated with the stripe density wave
order in some cuprates \cite{tranq} and nickelates, \cite{tranq1,cheong}
with the commensurate or incommensurate charge ordering in manganites.
\cite{cheong1}

The ordered phase generally evolves out of a metallic phase. In the weak
$el-ph$ coupling regime the transition is the well-understood $CDW$ 
instability of the Fermi liquid. \cite{gruner} 
The related equations predict a ratio of $T=0$ gap $\Delta$ to the ordering 
temperature $T_{CO}$ close to the $BCS$ value $\Delta/k_{B} T_{CO}=3.52$. 
However in quasi-one-dimensional materials this ratio assumes larger values 
because of critical fluctuations in low dimensionality. 
Furthermore values of the ratio larger than $10$ have been 
measured in quasi-two-dimensional and three-dimensional materials.
\cite{wilson,mcwhan,testardi,katsu,calvani,kim}
In $Ni$ compounds \cite{katsu} 
 $ \Delta/ k_{B} T_{CO} \simeq 13$ and in manganites \cite{kim}  
a ratio of order of $30$ have been deduced by the energy gap fitted by optical
measurements. 
This suggests that a strong coupling $CO$ phase develops in novel materials 
and that the lattice degrees of freedom can be important for stabilizing the
ordered state. \cite{jung} 
A very useful approach to investigate the properties of the ordered state is 
to study the optical absorption of $CO$ materials. 
\cite{gruner,jung,gruner1,degiorgi,vescoli}  
In particular the lattice fluctuation effects \cite{ross,degiorgi1,schwartz} 
and the response of the system in the strong coupling $CO$ regime 
\cite{katsu,jung,calvani1} have been studied.
Indeed the inverse square-root singularity expected for the case of a static 
distorted lattice can be removed by the fluctuation effects giving rise to 
a subgap tail absorption. \cite{degiorgi1}  
Furthermore the presence of a polaronic peak in the $MIR$ band of $Ni$ 
compounds observed above and below the transition temperature can be explained 
only within a theory valid for intermediate to strong coupling regime. 
\cite{katsu,jung} 

Theoretically one of the simplest models of electrons interacting with phonons
and developing a $CO$ state is the Holstein model. \cite{holstein}
The $CDW$ correlations and the transition temperature have been studied within
this model by Monte Carlo simulations 
\cite{marsiglio,noack,vekic,freericks,niyaz}
showing that the ordered phase is stable at half-filling.
Recently, using the dynamical mean-field theory ($DMFT$), the issue of the 
non-perturbative coupling scenario of the $CO$ phase and of the origin of the 
large ratio of the zero temperature gap to the transition temperature has been
addressed. \cite{ciuchi,millis,millis1}
The results point out that the scattering of electrons by the 
phonons involved in the lattice distortions represents the crucial effect in 
the coupling regime relevant to most $CDW$ materials.

In this paper we extend the variational scheme proposed for the spinless 
Holstein model at half-filling by H. Zheng, D. Feinberg and M. Avignon 
\cite{zheng} ($ZFA$) to finite temperature.
We note that at zero temperature the $ZFA$ results are in good agreement with 
different numerical works.\cite{hirsch,ross1,weisse} 
Actually this approach introduces lattice fluctuations on the Peierls 
dimerization since it takes into account the nonadiabatic polaron formation. 
Within this scheme all electrons in the Fermi sea are involved in the 
$el-ph$ scattering leading, above the $CO$ transition, to a stable 
phase of disordered small polarons ($SP$) in the intermediate to strong 
coupling regime. 
Thus, at half-filling the phase diagram is consistent with previous $DMFT$ 
results pointing out a strong dependence of the transition temperature on the 
adiabaticity particularly in the weak to intermediate coupling regime.
\cite{ciuchi,millis,millis1} 
Within the $ZFA$ scheme we have derived the spectral properties useful to 
characterize the different phases of the system.       
Next, by using the formalism of generalized Matsubara Green's functions
\cite{schna,loos1,loos2,kada,fehske},
we have determined the scattering rate of the quasi-particles finding
that, in the $CO$ phase, the single-phonon emission and absorption
represent the main mechanism of damping. 
The lattice fluctuation effects included in the scattering rate are
beyond the $ZFA$ scheme and modify the density of states that, this
way, is able to capture the features of the intermediate coupling
regime in agreement with previous $DMFT$ studies. \cite{ciuchi,millis,millis1}
 
The scattering rate turns out fundamental also to derive the 
optical properties of the system.
With decreasing $T$ or increasing the $el-ph$ coupling, our conductivity 
spectra are characterized by a 
transfer of spectral weight from low to high energies with the opening of 
an optical gap.
The effect of the lattice fluctuations is able to smooth the inverse 
square-root singularity of the mean-field approach in the optical absorption 
and induces a subgap tail.
In the weak to intermediate coupling regime, a two-peak structure is 
obtained within the $CO$ phase at low frequency: 
the remnant Drude-like term and the incipient absorption band centered around 
the gap energy.

In section I the model and the variational approach are reviewed; in section
II the spectral properties are discussed; in section III the damping of the 
particle motion is calculated; in section IV the optical properties are 
deduced.

\section{Model and variational approach} 
In this paper we study the spinless Holstein model at half-filling.
\cite{holstein} 
The Hamiltonian reads
 
\begin{eqnarray}
H=&&-t\sum_{<i,j>} 
c^{\dagger}_{i}c_{j}
 +\omega_0 \sum_{i}a^{\dagger}_{i}a_{i}
+g \omega_0 \sum_{i} c^{\dagger}_{i}c_{i} \left( a_{i}+a^{\dagger}_{i} 
\right)
  \nonumber \\
&& - \mu \sum_{i} c^{\dagger}_{i} c_{i} .  \label{1r}
\end{eqnarray}
Here  $t$ is the electron transfer integral between nearest neighbor ($nn$)
sites $<i,j>$,  
$c^{\dagger}_{i} \left( c_{i} \right)$ creates (destroys) an electron at the 
i-th site  and $\mu$ is the chemical potential.
In the second term of eq.(\ref{1r}) $ a^{\dagger}_{i} \left( a_{i} \right)$ 
is the creation (annihilation) phonon operator at the site i,
$\omega_0=\sqrt{k/M}$ denotes the frequency of the optical phonon mode, 
with $k$ restoring force per length unit and $M$ mass of the local 
oscillator.
The parameter $g$ represents the coupling constant between 
electrons and phonons 
\begin{equation}
g=\frac {A}{\omega_0 \sqrt{2 M \omega_0} },
  \label{1r1}
\end{equation}
where $A$ is the energy per displacement unit due to the coupling of the 
charge carriers with the lattice. The dimensionless parameter $\lambda$ 
\begin{equation}
\lambda= \frac{g^2 \omega_0}{2 D t},
  \label{1r1b}
\end{equation}
with $D$ dimensionality of the system, 
indicating the ratio between the $SP$ binding energy and the energy gain of 
an itinerant electron on a rigid lattice, is useful to measure the 
strength of the $el-ph$ interaction in the adiabatic regime.
We consider spinless electrons since they, even if at a very rough level, 
mimic the action of an on-site Coulomb repulsion preventing the formation of 
local pairs.

The hopping of electrons is supposed to take place between the equivalent 
$nn$ sites of a simple cubic lattice separated by the distance 
$|n-n^{\prime}|=a$. 
The units are such that the Planck constant $\hbar=1$, the Boltzmann constant
$k_B$=1 and the lattice parameter $a$=1.

Following the ZFA variational scheme, \cite{zheng}  
we perform three successive canonical transformations to treat the 
electron-phonon interaction variationally and to introduce the charge-ordering 
solution. 
The first is the variational Lang-Firsov unitary transformation \cite{zheng}
\begin{equation}
U_{1}=\exp \left[ g \sum_{j}  
\left( f c^{\dagger}_{j} c_{j} +\Delta_{j} \right) 
\left( a_{j}-a^{\dagger}_{j} \right)
\right],                                            
\label{2r}
\end{equation}
where $f$ and $\Delta_{j}$ are variational parameters.
The quantity  $f$ controls the degree of the polaron effect and $\Delta_{i}$
denotes a displacement field describing lattice distortions due to the average
electron motion.
The second transformation is 
\begin{equation}
U_{2}=\exp \left[\alpha \sum_{j} \left( a^{\dagger}_{j}
a^{\dagger}_{j} -a_{j}a_{j} \right)\right],          
\label{3r}
\end{equation}
where the variational parameter $\alpha$ determines a phonon frequency 
renormalization.
The transformed Hamiltonian 
$\tilde{H}=  U_{2}^{-1} U_{1}^{-1} H U_{1} U_{2} $ is

\begin{eqnarray}
\tilde{H}=&& 
-t \sum_{<i,j>} X^{\dagger}_i X_j  c^{\dagger}_{i} c_{j} 
+\bar{\omega}_0 \sum_{i} a^{\dagger}_{i}a_{i}+
L \omega_0 \sinh^2\left(2\alpha\right)
+  g^{2} \omega_{0} \sum_{i} \Delta_{i}^{2}    \nonumber \\
&& 
+\omega_0 \sinh\left(2 \alpha\right) \cosh\left( 2 \alpha\right) 
\sum_{i} \left( a^{\dagger}_{i}a^{\dagger}_{i} +a_{i}a_{i} \right)
-g \omega_{0} e^{2 \alpha}
\sum_{i} \Delta_{i} \left( a_{i} +a^{\dagger}_{i} \right) \nonumber \\
&& 
+g \omega_{0} \left( 1-f \right) e^{2 \alpha} 
\sum_{i} c^{\dagger}_{i}c_{i} \left( a_{i} +a^{\dagger}_{i} \right)
+  \sum_{i} c^{\dagger}_{i}c_{i} \left( \eta_{i} -\mu \right),
\label{4r}
\end{eqnarray}
where we have the phonon operator  $X_i$  

\begin{equation}
X_i=\exp \left[ g f e^{-2 \alpha} \left( a_{i} - a^{\dagger}_{i} \right)
\right],
\label{6r}
\end{equation}
the renormalized phonon frequency  
$ \bar{\omega}_0=\omega_0 \cosh(4 \alpha)$, the number of  lattice  
sites $L$
and the quantity $\eta_{i}$

\begin{equation}
\eta_{i} = g^{2} \omega_{0}  f \left( f-2 \right)+2 g^{2} \omega_{0} 
\left( f-1 \right) \Delta_{i}. 
\label{8r}
\end{equation}

At half-filling the charge-ordered solution is obtained by assuming
\begin{equation}
\Delta_{i}=\Delta+\Delta_{CO} e^{i \vec{Q} \cdot \vec{R_{i}} },
\label{8ra}
\end{equation}
where $\Delta$ represents the lattice distortion unaffected by the 
instantaneous position of electrons and $\Delta_{CO}$ the additional local 
lattice distortion due to the Peierls dimerization with 
$ \vec{Q}= (\pi,\pi,\pi)$.

In the ZFA approach the free energy is deduced employing the Bogoliubov 
inequality and introducing a test Hamiltonian characterized by non 
interacting electron and phonon degrees of freedom such that 
$ \langle \tilde{H}-H_{test} \rangle _{t}=0 $, where 
$ <>_t $ indicates a thermodynamic average obtained by using $H_{test}$.
The test Hamiltonian is given by

\begin{eqnarray}
H_{test} &=&
-t_{eff} \sum_{<i,j>} c^{\dagger}_{i}c_{j} +
\bar{\omega}_0 \sum_{i}a^{\dagger}_{i}a_{i}+
L \omega_0  \sinh^2\left( 2 \alpha\right)   
+L g^{2} \omega_{0}  \left( \Delta^{2}+\Delta^{2}_{CO} \right)   \nonumber \\
&& 
-2 g^{2} \omega_{0} \Delta_{CO} \left( 1-f \right) 
\sum_{i} e^{i \vec{Q} \cdot \vec{R_{i}} } c^{\dagger}_{i}c_{i}  
-\mu_{0} \sum_{i} c^{\dagger}_{i}c_{i},
\label{9r}
\end{eqnarray}
where the subsidiary chemical potential $\mu_{0}$ is 
\begin{equation}
\mu_{0} =\mu-g^{2} \omega_{0}  f \left( f-2 \right)+2 g^{2} \omega_{0} 
\left( f-1 \right) \Delta.
\end{equation}
The quantity $t_{eff}= t  e^{-S_{T}}$ denotes the effective transfer 
integral, where the quantity 
\begin{equation}
S_{T}=g^{2} f^{2} e^{-4 \alpha} \left( 2N_0+1 \right)
\end{equation}
controls the band renormalization due to the antiadiabatic polaron effect,
with  $N_0$ the average number of phonons with frequency $\bar{\omega}_0 $.
The test Hamiltonian is diagonalized by a third canonical Bogoliubov 
transformation \cite{zheng} yealding 
\begin{eqnarray}
\tilde{H}_{test} &=&
\sum_{{\bf{k}}\epsilon NZ} 
\left( \xi_{\bf{k}}^{(+)}  -\mu_{0} \right) d^{\dagger}_{\bf{k}} d_{\bf{k}} +
\sum_{{\bf{k}}\epsilon NZ} 
\left( \xi_{\bf{k}}^{(-)}  -\mu_{0} \right)  p^{\dagger}_{\bf{k}} p_{\bf{k}}+
\bar{\omega}_0 \sum_{\bf{q}}a^{\dagger}_{\bf{q}}a_{\bf{q}}+
\nonumber \\
&& 
+L \omega_0  \sinh^2\left( 2 \alpha\right)   
+L g^{2} \omega_{0} \left( \Delta^{2}+\Delta^{2}_{CO} \right), 
\label{9rb}
\end{eqnarray}
where $d^{\dagger}_{\bf{k}} \left( d_{\bf{k}} \right)$ creates (destroys) 
a quasi-particle in the  upper band 
$ \xi_{\bf{k}}^{(+)} = \sqrt{ \tilde{\epsilon}^{2}_{ \bf{k}}+E^{2} } $ ,
 $p^{\dagger}_{\bf{k}} \left( p_{\bf{k}} \right)$ creates (destroys) 
a quasi-particle in the lower band 
$ \xi_{\bf{k}}^{(-)} = -\sqrt{ \tilde{\epsilon}^{2}_{\bf{k}}+E^{2} } $
and $\tilde{\epsilon}_{\bf{k}}$ is the polaronic band.
We note that $NZ$ indicates the New Brillouin Zone defined by the condition
$\tilde{\epsilon}_{\bf{k}} \leq  0$.
In the charge-ordered phase a gap opens between the upper and lower bands in 
the quasi-particle spectrum and it is twice the quantity 
\begin{equation}
E= 2  g^{2} \omega_{0} \Delta_{CO} \left( 1-f \right) .
\end{equation}
At half-filling $(\rho=0.5)$, the variational free energy of the system is

\begin{eqnarray}
\frac{F}{L}=&& 
f^{el} + T \log{\left(1-e^{-\beta \bar{\omega}_0}\right)}+ 
\omega_0  \sinh^2\left( 2 \alpha\right) 
+ g^{2} \omega_{0} \left( \Delta^{2}+\Delta^{2}_{CO} \right)+
\mu(\rho) \rho   
\label{11r}
\end{eqnarray}
where $\beta$ is the inverse of the temperature.
The electron free energy reads 

\begin{equation}
f^{el} = T 
\int_{-\tilde{W}}^{0} d \epsilon g(\epsilon) 
\log\left[ n_{F}(\xi^{(+)}) n_{F}(\xi^{(-)})   \right],
\end{equation}
where $g \left( \epsilon \right)$ is the density of states, 
$\tilde{W}= 6 t_{eff}$ is the renormalized band half-width, 
$n_{F}(\epsilon)$ is the Fermi distribution, 
$\xi^{(+)}=\sqrt{\epsilon^2+E^2} $ and  
$\xi^{(-)}=-\xi^{(+)}$.
In order to simplify the calculations, in the three dimensional case 
we consider a semicircular density of states

\begin{equation}
g(\epsilon)=\left( \frac{2}{\pi \tilde{W}^2} \right) 
\theta (\tilde{W}-| \epsilon| )   
\sqrt{ \tilde{W}^2-\epsilon^2 },
\label{12r}
\end{equation}
where $\theta (x)$ is the Heaviside function. 
Actually $ g \left( \epsilon \right) $ represents a simple 
approximate expression for the exact density of states and it is 
generally used for a 3-D lattice. \cite{econo,georges} 

Within the variational approach the kinetic energy mean value 
$<\hat{T}>$ 
\begin{equation}
<\hat{T}>= \int_{- \tilde{W}}^{0} d \epsilon g(\epsilon)
\frac { \epsilon^{2} }{\sqrt{\epsilon^2+E^2 }  } 
\left[ n_{F}(\xi^{(+)}) -n_{F}(\xi^{(-)})   \right]
\label{562ra}
\end{equation}
and the electron order parameter $m_e$
\begin{equation}
m_e= \left( \frac{1}{L} \right)
\sum_{i} e^{i \vec{Q} \cdot \vec{R_{i}} } 
\langle c^{\dagger}_{i}c_{i} \rangle = 
E \int_{-\tilde{W}}^{0} d \epsilon \frac { g(\epsilon) } 
{ \sqrt{\epsilon^2+E^2 }  } 
\left[ n_{F}(\xi^{(-)}) -n_{F}(\xi^{(+)})   \right]
\label{562rb}
\end{equation}
can be evaluated. The $CO$ phase is characterized by the order 
parameter different from zero.

In this paper we discuss results valid in three dimensions. In Fig.1 we 
report the phase diagram obtained within our approach for $W= 6 \omega_{0}$,
with $W$ the bare band half-width. We compare it with the mean-field result 
obtained in the adiabatic limit for $f=0$ and $\alpha=0$ that overestimates 
the transition temperature. The $A$ phase represents the Fermi-liquid-like 
normal state, while the $B$ phase the disordered localized $SP$ normal state.
This latter phase is determined when there is absence of order 
($\Delta_{CO}=0$) and the lattice presents the largest distortions ($f=1$). 
While the transition from $A$ to $CO$ is continuous, the transition from $CO$
to $B$ and the crossover from $A$ to $B$ is rather discontinuous. We stress 
that a discontinuous character for the transition can be considered as a 
drawback of the $ZFA$ approach. \cite{zheng} 
However, the phase diagram bears a strong resemblance to that derived with 
more sophisticated tecniques. \cite{ciuchi,millis}
   
In the inset of Fig.1, the ratio of $T=0$ gap $\Delta$ to the ordering 
temperature $T_{CO}$ is shown. It is interesting to note that within the 
$CO$ phase it is possible to distinguish two regimes. The first is the 
mean-field regime in the sense that the ratio assumes values near the $BCS$ 
$\Delta/k_{B} T_{CO}=3.52$, while the second regime exhibits larger values of 
this quantity.
For instance at $\lambda=0.9$ the ratio $\Delta/k_{B} T_{CO}$ is about $13$, 
the value measured in nickelates. \cite{katsu}
Hence, a strong coupling $CO$ phase develops starting from intermediate
values of the $el-ph$ coupling. This $CO$ regime is characterized by a gap
of the order of $2 g^{2} \omega_{0} $ and by distortions comparable to those
of the disordered $SP$ phase. 

Previous $DMFT$ results \cite{millis1} have pointed out a strong dependence 
of the transition temperature on the adiabaticity particularly in the weak to
intermediate coupling regime. 
Therefore we have determined the transition temperature for two different 
values of the adiabaticity ratio (see Fig.2a ). 
The temperature $T_{CO}$ strongly depends on the adiabaticity  for 
intermediate $el-ph$ couplings. 
Here the lattice fluctuations are very effective in reducing $T_{CO}$.  
The variation $\Delta T_{CO}=T_{CO}^{*}-T_{CO}$ of the transition 
temperatures due to different masses of the oscillators is reported in
Fig.2b. The ratio between the old and the new masses simulates the 
isotopic substitution of the oxygen from $^{16} O$ ($M$) to $^{18} O$ 
($M^*$).
Actually in many perovskite oxides the presence of the oxygen atoms induces 
the $el-ph$ coupling. \cite{imada} 
Therefore we can consider a change of the values of $\omega_0$ and $g$ 
to $\omega_0^*=\omega_0\sqrt{M/M^*}$ and $g^*=g\left(M^*/M\right)^{1/4}$ 
respectively. \cite{perroni0} 
The softening of the phonon frequency induces an increase of the transition 
temperature. Furthemore, as a function of $\lambda$, the difference between 
the two ordering temperatures is characterized by a maximum in the region of
weak to intermediate $el-ph$ couplings.

\section{Spectral properties}

In this section we calculate the spectral properties within the $ZFA$ 
approach. They are discussed in order to characterize the different phases
of the diagram reported in Fig.1. 

The electron Matsubara Green's function can be disentangled into electronic 
and phononic terms \cite{perroni} by using the test Hamiltonian (\ref{9r}), 
hence    

\begin{equation}
{\mathcal G} \left( \bf {k},\tau \right)=
- \left( \frac{1}{L} \right) \sum_{i,j} 
e^{ i \bf{k} \cdot \left( \bf{R}_i - \bf{R}_j \right) }  
\langle T_{\tau} \bar{c}_i \left( \tau \right) c_j^{\dagger} \rangle_t 
 \langle T_{\tau} \bar{X}_i \left( \tau \right) X_j^{\dagger} \rangle_t ,
\label{16ar}
\end{equation}
where we have 
$\bar{c}_i \left( \tau \right)= e^{\tau H_t} c_i e^{-\tau H_t}$ and
$\bar{X}_i \left( \tau \right)= e^{\tau H_t} X_i e^{-\tau H_t}$.
The Green's function in Matsubara frequencies $\omega_n$ becomes

\begin{eqnarray}
{ \mathcal G} \left( {\bf k},i \omega_n \right)= &&  
 e^{-S_T} { \mathcal G}^{(c)} \left( {\bf k},i \omega_n \right)+ 
\nonumber \\
&& e^{ -S_T } \left( \frac{1}{N} \right)   \sum_{\bf{k}_1}
\left( \frac{1}{\beta} \right) \sum_{m} 
{\mathcal G}^{(c)} \left( {\bf k}_1,i \omega_m \right)
\int_0^{\beta} e^{(i \omega_n -i \omega_m ) \tau}
\{ e^{s\cosh\left[\bar{\omega}_0\left(\tau-\frac{\beta}{2}\right)\right]}
-1 \},
\label{17r}
\end{eqnarray}
where $s$ is
\begin{equation}
s=2 f^2 g^2 e^{-4 \alpha} \left[ N_0 \left( N_0+1 \right) \right].
\label{17as}
\end{equation}
In eq.(\ref{17r}) ${ \mathcal G}^{(c)} \left( {\bf k},i \omega_n \right)$ is 
the Green's function 
\begin{equation}
{ \mathcal G}^{(c)} \left( {\bf k},i \omega_n \right)=
u^2_{\bf{k}}  { \mathcal G}^{(+)} \left( {\bf k},i \omega_n \right)+
v^2_{\bf{k}}  { \mathcal G}^{(-)} \left( {\bf k},i \omega_n \right),
\label{17ra}
\end{equation}
where ${ \mathcal G}^{(+)} \left( {\bf k},i \omega_n \right)$ and 
${ \mathcal G}^{(-)} \left( {\bf k},i \omega_n \right)$ are the  
Green's functions associated to the quasi-particles of the upper and lower 
bands, with $u^2_{\bf{k}}$ given by

\begin{equation}
u^2_{\bf{k}}=
\frac {1} {2}
\left[ 
1+ \frac {\tilde{\epsilon}_{\bf{k}} } 
{ \sqrt{\tilde{\epsilon}_{\bf{k}}^2+E^2 }  } 
\right]
\label{17rb}
\end{equation}
and  $v^2_{\bf{k}}$ by

\begin{equation}
v^2_{\bf{k}}=
\frac {1} {2}
\left[ 
1 - \frac {\tilde{\epsilon}_{\bf{k}} } 
{ \sqrt{\tilde{\epsilon}_{\bf{k}}^2+E^2 }  } 
\right].
\label{17rc}
\end{equation}
Two physically distinct terms \cite{perroni} appear in the Eq.(\ref{17r}).
The first derives from the coherent motion of quasi-particles and their 
surrounding phonon cloud, while the incoherent term is due to processes
changing  the number of phonons during the hopping of the charges.

Making the analytic continuation $ i \omega_n \rightarrow \omega +i \delta $,
one obtains the retarted Green's function $ G_{ret} ( {\bf k}, \omega)$ and 
the spectral function 

\begin{eqnarray}
A( {\bf k}, \omega) =&& -2 \Im G_{ret} ( {\bf k}, \omega) =
2 \pi e^{-S_T} 
\left[
u^2_{\bf{k}} \delta \left( \omega - \xi_{{\bf k}}^{(+)} \right)+
v^2_{\bf{}k} \delta \left( \omega - \xi_{{\bf k}}^{(+)} \right)
\right]+
\nonumber \\
&&  2 \pi e^{-S_T}  \sum_{l(\neq 0)- \infty}^{+ \infty}
I_l (s) e^{ \frac{ \beta l \bar{\omega}_0}{2}}
[ 1 - n_F(\omega - l \bar{\omega}_0)] K(\omega - l \bar{\omega}_0) +
\nonumber \\
&& 2 \pi e^{-S_T}  \sum_{l(\neq 0)- \infty}^{+ \infty} I_l (s) 
e^{ \frac{ \beta l \bar{\omega}_0}{2}}
n_F(\omega + l \bar{\omega}_0) 
K(\omega+ l \bar{\omega}_0)+ 
\nonumber \\
&& 2 \pi e^{-S_T} [I_0(s) -1 ] K(\omega),
\label{21r}
\end{eqnarray}
where the function $K(\omega)$ is
\begin{equation}
K(\omega)=H(\omega)+H(-\omega),
\label{16rd}
\end{equation}
with
\begin{equation}
H(\omega)=\frac { g\left( \sqrt{\omega^2-E^2} \right) }
{\sqrt{1-\frac{E^2}{\omega^2}}}
\theta \left(\omega-E \right) \theta \left(\sqrt{E^2+W^2}-\omega \right).
\label{16re}
\end{equation}
In the normal phase the first term of eq.(\ref{21r}) represents the purely 
polaronic band contribution and shows a delta behavior. 
In the $CO$ phase, this term is equal to the result of the mean-field 
approach when one neglects the renormalization of the upper and lower band 
due to the polaron effect.
The remaining terms of eq.(\ref{21r}) provide the incoherent contribution 
that spreads over a wide energy range. 

Thus we get the renormalized density of states $ N( \omega)$ that is 
normalized to unity.
In Fig.3a we report the density of states for three different values of the 
$el-ph$ coupling at $T=0.023 W$. It is apparent that in the normal phase 
($\lambda=0.15$) the weight of the coherent term is prevalent and the 
density of states is practically unchanged with respect to the non-interacting 
case.
Entering the $CO$ phase, a gap opens in the quasi-particle spectrum and it 
broadens with increasing the parameter $\lambda$. We note that, at the 
energies corresponding to the gap, the inverse square-root singularity 
occurs. The other sharp features of the density of states are due to 
one-phonon and two-phonon processes in the upper and lower bands.

At higher temperatures (Fig. 3b) similar features are found.
However, in this case, for large $el-ph$ couplings, the density of states is
determined by the incoherent dynamics of the small polaron excitations.
\cite{perroni} 
It is made up of two bands peaked approximatively around $-g^2 \omega_0$ and
$g^2 \omega_0$, whose heights are equal at half-filling. We stress 
the evolution of the gap energy of the $CO$ phase toward the characteristic 
energy $2 g^{2} \omega_{0} $ that represents the energy range between the 
two peaks of the $SP$ density of states.

\section{Damping}

In this section we deal with the $el-ph$ self-energy that includes lattice
fluctuation effects beyond the $ZFA$ approach. This quantity allows 
to determine the scattering rate of the quasi-particles of the upper and 
lower bands and a better density of states. This analysis will play an
essential role in the infrared absorption calculations that is the
main aim of this paper. For sake of clarity we will put forward this 
discussion. 

Retaining only the dominant autocorrelation terms at the second step of 
iteration \cite{schna,loos1,loos2,kada,fehske}, we derive the local 
self-energy

\begin{eqnarray}
\Sigma^{(2)} \left(i \omega_n \right)=  
\left( \frac{1}{N} \right) \sum_{\bf{k}_1}
\left( \frac{1}{\beta} \right) \sum_{m}
{\mathcal G}^{(c)}  \left( {\bf k}_1,i \omega_m \right)
\int_0^{\beta} d\tau e^{(i \omega_n -i \omega_m ) \tau}
[f_1(\tau)+f_2(\tau)],
\label{34r}
\end{eqnarray}
where $f_1(\tau)$ is

\begin{eqnarray}
f_1(\tau)=Z t^2 e^{ -2 S_T} 
\left[
e^{ z\cosh \left[\bar{\omega}_0\left(\tau-\frac{\beta}{2}\right)\right] }- 1 
\right],
\label{35r}
\end{eqnarray}
$z=2s$, with $s$ given by eq.(\ref{17as}), $Z$ the number of $nn$
sites, and $f_2(\tau)$ is 

\begin{eqnarray}
f_2(\tau)=2 g^2 \omega_0^2 e^{ 4 \alpha } (1-f)^2 
[N_0(N_0+1)]^{ \frac{1}{2}} 
\cosh 
\left[ 
\bar{\omega}_0 \left( \tau- \frac{\beta}{2} \right) 
\right]. 
\label{37r}
\end{eqnarray}
Other terms beyond the autocorrelation ones induce a dependence of the
self-energy on the momentum and could be included. \cite{fehske}
We note that the fundamental autocorrelation
contribution presents some analogies with the corresponding
self-energy within $DMFT$ approaches where the independence on the
momentum is the main assumption when the interaction term is local.

Making the analytic continuation $i \omega_n \rightarrow \omega +i\delta$, 
we can calculate the scattering rate of the quasi-particles of the upper 
band 
	
\begin{equation}
\Gamma_{+}( {\bf k})=\tilde{\Gamma} ( \omega = \xi^{(+)}_{{\bf k}}  )
=-2 \Im \Sigma^{(2)}_{ret} 
\left( \omega= \xi^{(+)}_{{\bf k}}  \right).
\label{42r}
\end{equation}

We can express the scattering rate in the following way 
\begin{equation}
\Gamma_{+}( {\bf k})=\Gamma_{+}(\xi^{(+)}_{{\bf k}})= 
\Gamma_{1-phon}(\xi^{(+)}_{{\bf k}})+ \Gamma_{multi-phon}(\xi^{(+)}_{{\bf k}})
 + \Gamma_{res}(\xi^{(+)}_{{\bf k}}),
\label{44r}
\end{equation}
where $\Gamma_{1-phon}$ is the contribution due to single phonon
processes only 

\begin{eqnarray}
\Gamma_{1-phon}(\xi^{(+)}_{{\bf k}})=&&
2 Z t^2 e^{ -2 S_T} 
I_1(z) 
\sinh \left( \frac{\beta \bar{\omega}_0 }{2} \right)
g_{1,l=1}(\xi^{(+)}_{{\bf k}})+
\nonumber \\
&& g^2 \omega_0^2 e^{ 4 \alpha } (1-f)^2 g_2(\xi^{(+)}_{{\bf k}}),
\label{45r}
\end{eqnarray}
$\Gamma_{multi-phon}$ represents the scattering rate by 
multiphonon processes 

\begin{equation}
\Gamma_{multi-phon}(\xi^{(+)}_{{\bf k}})=
2 Z t^2 e^{ -2 S_T} 
\sum_{l=2}^{+ \infty} I_l(z) 
\sinh \left( \frac{\beta \bar{\omega}_0 l}{2} \right)
g_{1,l}(\xi^{(+)}_{{\bf k}})
\label{46r}
\end{equation}
and $\Gamma_{res}$ denotes a residue term due to the difference between the 
coherent and the incoherent contribution \cite{mahan}
\begin{equation}
\Gamma_{res}(\xi^{(+)}_{{\bf k}})=
Z t^2 e^{ -2 S_T} 
\left[ 
I_0(z)-1 
\right]
B(\xi^{(+)}_{{\bf k}}).
\label{47r}
\end{equation}
In the previous equations the function $g_{1,l}(\xi^{(+)}_{{\bf k}})$ reads 

\begin{eqnarray}
g_{1,l}(\xi^{(+)}_{{\bf k}})
& = & 
\left[ 
N_0 (l \bar{\omega}_0 )+n_F(\xi^{(+)}_{{\bf k}} +l \bar{\omega}_0)
\right]
B(\xi^{(+)}_{{\bf k}} +l \bar{\omega}_0 )  \nonumber \\
& + & 
\left[ 
N_0 (l \bar{\omega}_0 )+1-n_F(\xi^{(+)}_{{\bf k}} -l \bar{\omega}_0)  
\right]
B(\xi^{(+)}_{{\bf k}} -l \bar{\omega}_0 )
\end{eqnarray}
and $g_2(\xi^{(+)}_{{\bf k}}) $
 
\begin{equation}
g_2(\xi^{(+)}_{{\bf k}})=
\left[
N_0 +n_F(\xi^{(+)}_{{\bf k}} +\bar{\omega}_0)
\right]
B(\xi^{(+)}_{{\bf k}} +\bar{\omega}_0 )+
\left[
N_0 +1-n_F(\xi^{(+)}_{{\bf k}} -\bar{\omega}_0)  
\right]
B(\xi^{(+)}_{{\bf k}} - \bar{\omega}_0 ).
\label{49r}
\end{equation}
Moreover the function $B(x)$ is equal to $2 \pi  K(x)$, with $K(x)$ given
in eq.(\ref{16rd}).
The decomposition of the scattering rate in three distinct terms
has been introduced in order to simplify the analysis of the results.

We define the scattering rate for the quasi-particles of the lower band 

\begin{equation}
\Gamma_{-}( {\bf k})=\tilde{\Gamma} ( \omega = \xi^{(-)}_{{\bf k}}  )
=-2 \Im \Sigma^{(2)}_{ret} 
\left(\omega= \xi^{(-)}_{{\bf k}}  \right),
\label{42rs}
\end{equation}
that turns out to be  equal to $\Gamma_{+}( {\bf k})$. Thus we can take into 
account only one scattering rate 
\begin{equation}
\Gamma( {\bf k})=\Gamma_{+}( {\bf k})=\Gamma_{-}( {\bf k}).
\end{equation}

In Fig.4a we report the scattering rate $\Gamma$ for different $el-ph$ 
couplings at low temperature. 
In the normal phase ($\lambda=0.15$) the scattering rate is zero within 
$\omega_0$ of the chemical potential $\mu$. Therefore at this
temperature the main mechanism of energy loss is the spontaneous
emission of phonons with frequency $\omega_0$ by the quasi-particles
(one phonon processes are prevalent). Thus the behavior of the 
scattering rate is determined by of the Fermi statistics since the 
quasi-particle excitations within $\omega_0$ of the Fermi energy 
cannot lose energy. \cite{perroni,mahan} 
In the $CO$ phase ($\lambda=0.50$) the gap in the scattering rate
opens at energies given by the sum of the gap energy 
(indicated by the arrows in the figure) and 
the phonon frequency. Since there are not available states within the gap,
only the quasi-particles with such energies can be scattered by the residue
$el-ph$ interaction.

In Fig.4b we concentrate on the scattering rate at a fixed $el-ph$ coupling
($\lambda=0.4$) for different temperatures. At low $T$ ($T=0.01 W$)
there is the gap due to the Peierls dimerization and the processes of 
phonon spontaneous emission by the quasi-particles. With rising 
temperature, we can follow the increase of the quantity $\Gamma$ in the 
normal phase because of the absorption and emission of phonons. In the
normal phase at finite temperatures, these scattering processes are
effective also in the energy range around the chemical potential where
there are able to enhance the rate.
Actually, at higher temperatures ($T=0.14 W$), the rate is weakly
dependent on the energies involved in the phonon scattering.
It is confirmed that in this regime of temperatures
the single-phonon emission and absorption give an important contribution
to the quantity $\Gamma$.

For large $el-ph$ couplings and for small polaron excitations, the quantity 
$\Gamma$ decreases when $T$ increases but it is always larger 
than $t_{eff}$, therefore the electronic states lose their individual 
characteristics and the electron motion is predominantly a diffusive process.
A high-temperature expansion \cite{loos1,loos2,perroni} provides the
scattering rate 

\begin{equation}
\Gamma= Z t^2 \sqrt{ \frac{\pi \beta}{\bar{z}} } e^{-\beta \bar{z}/4 } 
=Z t^2 \sqrt{ \frac{\pi \beta}{4 E_a} } e^{-\beta E_a }, 
\label{42rsw}
\end{equation}
with $\bar{z}=2 g^2 \omega_0$ and $E_a = \bar{z}/4$. This value
coincides with the well-known rate of the Holstein polaron in the 
``classical'' limit, with $E_a$ typical activation energy for hopping.
 \cite{holstein}

The introduction of the damping allows to improve the approximations of 
calculation for the spectral properties. This can be carried out substituting 
in eq.(\ref{17r}) the new Green's function $\tilde{ {\mathcal G}}^{(c)} $ 
\begin{equation}
\tilde {{ \mathcal G}}^{(c)} \left( {\bf k},i \omega_n \right)=
u^2_{\bf{k}} \tilde{ { \mathcal G}}^{(+)} \left( {\bf k},i \omega_n \right)+
v^2_{\bf{k}}  \tilde {{ \mathcal G}}^{(-)} \left( {\bf k},i \omega_n \right).
\end{equation}
for ${\mathcal G}^{(c)}$.
The Green's function 
$\tilde{ {\mathcal G}}^{(\nu)}   \left( {\bf k},i \omega_n \right)   $ is
expressed as 

\begin{equation}
\tilde{ {\mathcal G} }^{(\nu)} ( {\bf k},i \omega_n )=
\int_{- \infty}^{+ \infty} \frac{ d \omega } {2 \pi} 
\frac
{ \tilde{ A }^{(\nu)}( {\bf k}, \omega )}
{i \omega_n - \omega }  ,
\label{38r}
\end{equation} 
where the spectral function $\tilde{ A }^{(\nu)}$ is assumed to be
\begin{equation}
\tilde{ A }^{(\nu)}( {\bf k}, \omega )=
\frac
{\Gamma( {\bf k})}
{ [\Gamma( {\bf k})]^2/4 +(\omega - \xi_{{\bf k}}^{(\nu)} )^2   },
\label{43r}
\end{equation}
with $\nu$ standing for $+$ or $-$.
We can determine a new electron spectral function and density of
states $N_{Fluct}$ that includes fluctuation effects beyond the $ZFA$ 
approach. 
These effects are able to change the density of states in the 
Fermi-liquid-like phase.
Indeed in this phase it is interesting to study the behavior of the
density as function of the temperature at a fixed value of the
coupling constant $\lambda$ (see Fig.5a). While the density obtained
within the $ZFA$ approach is similar to the bare one (solid line in
figure), now $N_{Fluct}$ shows an enhancement at the Fermi energy in
the low temperature regime ($T=0.07 W$) and a depression at the same 
energy in the high $T$ case ($T=0.20 W$). 
Actually at low $T$ the processes of phonon spontaneous
emission subtract spectral weight to states of energy larger than
$\omega_0$ compared with the chemical potential inducing an increase at
the Fermi energies where the states are not damped. At intermediate
temperatures ($T=0.12 W$), where real phonons are present, the absorption
and emission processes are effective also around the chemical potential
and oppose this enhancement, so that the density of states at the Fermi
energy is nearly unchanged. At higher temperatures these processes
dominate causing a reduction of the spectral weight at the chemical
potential.

In Fig.5b the density of states $N_{Fluct}$ is plotted as a function
of the coupling constant $\lambda$ showing that it is
able to capture the features of the intermediate coupling regime. For
these values of parameters, the density within the $ZFA$ approach
shows a discontinuous transition from a lightly renormalized function 
to a strongly modified density with a marked reduction at the chemical
potential. Due to the included lattice fluctuations, $N_{Fluct}$
evolves gradually from the bare density to the characteristic small 
polaron function showing pseudogap features in agreement with previous
studies. \cite{ciuchi,millis,millis1}
Not only in the normal phase but also in the $CO$ state, the quantity 
$\Gamma$ changes the spectral properties. Actually, in the boundary
phase region below the transition temperature of the $ZFA$ approach, a
pseudogap opens in the density of states as a precursor effect of a
gap at lower temperatures and stronger $el-ph$ couplings.

Thus the lattice fluctuation effects introduced by means a self-energy
insertion turn out to be able to correct the drawbacks of the $ZFA$ approach.
Next we will focus on the optical properties analyzing the prominent 
role of the lattice fluctuations in determining the infrared absorption of 
the system.

\section{Optical properties}

In this section we deal with the optical properties near and within
the $CO$ phase.
Calculating the real part of the conductivity in a regime of linear response
reduces to evaluate the retarded current-current correlation function
\begin{equation}
\Re \sigma_{\alpha,\beta}(\omega)= 
- \frac{ \Im \Pi_{\alpha,\beta}^{ret}(\omega) }{\omega}.
\label{54r}
\end{equation}
The electron motion will be supposed to take place between the 
equivalent $nn$ sites of the cubic lattice, hence the tensor 
$\sigma_{\alpha,\beta}$ is assumed to be diagonal with mutually equal elements 
$\sigma_{\alpha,\alpha}$.

Performing the two canonical transformations (\ref{2r},\ref{3r}) and 
making the decoupling\cite{perroni} of the correlation function in the 
electron and phonon terms through the introduction of $H_{test}$ (\ref{9r}), 
we get in Matsubara frequencies

\begin{equation}
\Pi_{\alpha,\alpha}(i \omega_n)=e^2 t^2 
\left(- \frac{1}{L} \right) \sum_{i,\delta} \sum_{i^{\prime},\delta^{\prime}}
( \delta \cdot \delta^{\prime} )
\int_0 ^{\beta} d\tau e^{i \omega_n \tau} 
\Phi (i,i^{\prime},\delta,\delta^{\prime},\tau)
\Delta (i,i^{\prime},\delta,\delta^{\prime},\tau),
\label{57r}
\end{equation}
where the function $\Delta$ denotes the electron correlation function
\begin{equation}
\Delta\left( i,i^{\prime},\delta,\delta^{\prime}, \tau \right)=
\langle T_{\tau} \bar{c}_i^{\dagger}(\tau) \bar{c}_{i+\delta \hat{\alpha}}
(\tau)
c_{i^{\prime}+\delta^{\prime} \hat{\alpha}}^{\dagger} c_{i^{\prime}} \rangle_t
\end{equation}
and the function $\Phi$ the phonon correlation function 

\begin{equation}
\Phi\left( i,i^{\prime},\delta,\delta^{\prime} ,\tau \right) =
\langle T_{\tau} \bar{X}_i^{\dagger}(\tau) \bar{X}_{i+\delta \hat{\alpha}}
(\tau)
X_{i^{\prime}+\delta^{\prime} \hat{\alpha}}^{\dagger} X_{i^{\prime}} \rangle_t.
\label{59r}
\end{equation}

To derive the optical properties, the role of the damping $\Gamma$ of the 
particle motion is fundamental. Since the electron correlation function can be 
expressed as a function of the Green's functions   $ {\cal G}^{(\nu)}$, 
the effect of the damping \cite{perroni} can enter our calculation 
substituting $ {\cal G}^{(\nu)}$ for $ \tilde{{\cal G}}^{(\nu)}$ given 
by eq.(\ref{38r}).
Furthermore, in order to simplify the analysis of our results, we
separate $\Phi$ into two contributions
\begin{eqnarray}
\Phi\left( i,i^{\prime},\delta,\delta^{\prime} \right) & = &
\left[ \langle X_i^{\dagger} X_{i+\delta \hat{\alpha}}\rangle_t \right]^2
\left\{
\Phi\left( i,i^{\prime},\delta,\delta^{\prime} \right) -
\left[ \langle X_i^{\dagger} X_{i+\delta \hat{\alpha}}\rangle_t \right]^2
\right\} \nonumber \\
&=&
 e^{ -2 S_T} +
\left[
\Phi\left( i,i^{\prime},\delta,\delta^{\prime} \right)- e^{ -2 S_T}
\right].
\label{63r}
\end{eqnarray}
Considering the two terms of eq.(\ref{63r}), the current-current correlation 
function can be 
written as

\begin{eqnarray}
\Pi_{\alpha,\alpha}(i \omega_n)=
\Pi^{(1)}_{\alpha,\alpha}(i \omega_n)+
\Pi^{(2)}_{\alpha,\alpha}(i \omega_n).
\label{65r}
\end{eqnarray}
The first term reads
\begin{equation}
\Pi^{(1)}_{\alpha,\alpha}(i \omega_n)=
 4 e^2 t^2 e^{ -2 S_T} 
 \left( \frac{1}{L} \right) \sum_{{\bf k}\epsilon NZ} \sin^2(k_{\alpha})
 \sum_{\nu_1,\nu_2}
h^{(\nu_1,\nu_2)} \left( {\bf k} \right)
S^{(\nu_1,\nu_2)}({\bf k},i \omega_{n}),
\end{equation}
where we have
\begin{equation}
h^{(+,+)} \left( {\bf k} \right)=
h^{(-,-)} \left( {\bf k} \right)=
(u^2_{\bf{k}} - v^2_{\bf{k}}     ) ^2,
\end{equation}

\begin{equation}
h^{(+,-)} \left( {\bf k} \right)=
h^{(-,+)} \left( {\bf k} \right)=
4 u^2_{\bf{k}} v^2_{\bf{k}}
\end{equation}
with $ S^{(\nu_1,\nu_2)}({\bf{ k}},i \omega_{n} )$ given by

\begin{equation}
S^{(\nu_1,\nu_2)}({\bf k},i \omega_{n})=\int_0 ^{\beta} 
d \tau e^{i \omega_n \tau}
\tilde{{\cal G}}^{(\nu_1)} ({\bf k},-\tau)  
\tilde{{\cal G}}^{(\nu_2)} ({\bf k},\tau).
\end{equation}
The second term of the current-current correlation function is obtained 
retaining only the main autocorrelation term 
$i=i^{\prime}$ and $ \delta=\delta^{\prime}$ 

\begin{equation}
\Pi^{(2)}_{\alpha,\alpha}(i \omega_n)=
\left( \frac {2 e^2 } {Z} \right)  
\left( \frac{1}{L^2} \right) 
\sum_{{\bf k} \epsilon NZ} 
\sum_{{\bf k}_1 \epsilon NZ}
\sum_{\nu_1,\nu_2}
p^{(\nu_1,\nu_2)} \left( {\bf k}, {\bf k}_1 \right)
T^{(\nu_1,\nu_2)}({\bf k}, {\bf k}_1 ,i \omega_{n}) ,
\label{67r}
\end{equation}
where we have
\begin{equation}
p^{(+,+)} \left( {\bf k},{\bf k}_1 \right)=
p^{(-,-)} \left( {\bf k},{\bf k}_1 \right)=
\left( u_{\bf{k}} u_{\bf{k}_1} - v_{\bf{k}} v_{\bf{k}_1}     \right) ^2,
\end{equation}

\begin{equation}
p^{(+,-)} \left( {\bf k},{\bf k}_1 \right)=
p^{(-,+)} \left( {\bf k},{\bf k}_1 \right)=
\left( u_{\bf{k}} v_{\bf{k}_1} + u_{\bf{k}_1} v_{\bf{k}} \right)^2,
\end{equation}
the function $ T^{(\nu_1,\nu_2)}({\bf{ k}},{\bf k}_1,i \omega_{n} )$ 

\begin{equation}
T^{(\nu_1,\nu_2)}({\bf k},{\bf k}_1,i \omega_{n})
=\int_0 ^{\beta} d \tau e^{i \omega_n \tau}
\tilde{{\cal G}}^{(\nu_1)} ({\bf k},-\tau)  
\tilde{{\cal G}}^{(\nu_2)} ({\bf k}_1,\tau)
f_1(\tau),  
\end{equation}
with $f_1(\tau)$ given by eq.(\ref{35r}).

We perform the analytic continuation $i \omega_n \rightarrow \omega +i\delta$,
and, clearly, the conductivity can be expressed as a sum of two terms 
\cite{perroni}

\begin{equation}
\Re \sigma_{\alpha,\alpha}(\omega)= 
- \frac{ \Im \left[ \Pi_{\alpha,\alpha}^{ret (1)}(\omega)+ 
\Pi_{\alpha,\alpha}^{ret (2)}(\omega) \right]} {\omega}
=\Re \sigma^{(coh)}_{\alpha,\alpha}(\omega)
+ \Re \sigma^{(incoh)}_{\alpha,\alpha}(\omega).
\label{68r}
\end{equation}
As in the spectral properties, the appearance of two physically distinct 
contributions, the coherent and incoherent one, occurs.
Actually the first term $\Re \sigma^{(coh)}_{\alpha,\alpha}$ is due to the 
charge transfer affected by the interactions with the lattice but not 
accompanied by processes changing the number of phonons. 
On the other hand, the incoherent term $ \Re \sigma^{(incoh)}_{\alpha,\alpha}$
in eq. (\ref{68r}) derives from inelastic scattering processes of emission 
and absorption of phonons.
The coherent conductivity is derived as 

\begin{equation}
\Re \sigma^{(coh)}_{\alpha,\alpha}(\omega)= 
\left( \frac{ 4 e^2 t^2}{\omega} \right) e^{ -2 S_T} 
\sum_{\nu_1,\nu_2}
\int_{-\tilde{W} }^{0} d \epsilon 
[n_F(\xi^{(\nu_1)}-\omega)-n_F(\xi^{(\nu_1)})]
\tilde{C}^{(\nu_1,\nu_2)}(\epsilon,\omega) 
h(\epsilon)
A^{(\nu_1,\nu_2)}(\epsilon),
\label{130ar}
\end{equation} 
where $\tilde{C}^{(\nu_1,\nu_2)}(\epsilon,\omega)$ is    

\begin{equation}
\tilde{C}^{(\nu_1,\nu_2)}(\epsilon,\omega)=  
\frac{ \Gamma(\epsilon) }
{ \Gamma^2(\epsilon)+
\left( \xi^{(\nu_2)} -\xi^{(\nu_1)} + \omega   \right)^2 }
\label{131ar}
\end{equation} 
and $h(\epsilon)$ is defined as

\begin{equation}
h(\epsilon)=\left( \frac{1}{N} \right) \sum_{{\bf k}} \sin^2(k_{\alpha})
\delta( \epsilon - \tilde{\epsilon}_{{\bf k}} ).
\label{132ar}
\end{equation} 
In eq.(\ref{130ar}) the function $A^{(\nu_1,\nu_2)}(\epsilon)$ is expressed by

\begin{equation}
A^{(+,+)}(\epsilon)=A^{(-,-)}(\epsilon)=
\frac{\epsilon^2}{\epsilon^2+E^2}
\end{equation}
and

\begin{equation}
A^{(+,-)}(\epsilon)=A^{(-,+)}(\epsilon)=
\frac{E^2}{\epsilon^2+E^2}.
\end{equation}
The latter term of the conductivity becomes

\begin{eqnarray}
\Re \sigma^{(incoh)}_{\alpha,\alpha}(\omega)= && 
\left( \frac{ 2 e^2 t^2}{\omega} \right) e^{ -2 S_T} 
\sum_{\nu_1,\nu_2}
\int_{-\tilde{W} }^{0} d \epsilon
\int_{-\tilde{W} }^{0} d \epsilon_1  
g(\epsilon) g(\epsilon_1) R^{(\nu_1,\nu_2)}(\epsilon,\epsilon_1,\omega)+
\nonumber \\
&& \left( \frac{ 2 e^2 t^2}{\omega} \right) e^{ -2 S_T} 
\left[ I_0(z)-1 \right]
\sum_{\nu_1,\nu_2}
\times
\nonumber \\
&& \times
\int_{-\tilde{W} }^{0} d \epsilon
\int_{-\tilde{W} }^{0} d \epsilon_1  
g(\epsilon) g(\epsilon_1)  
\left[ n_F(\xi^{(\nu_2)}_1-\omega)-n_F(\xi^{(\nu_2)}_1) \right]
C^{(\nu_1,\nu_2)}(\epsilon,\epsilon_1,\omega), 
\label{134ar}
\end{eqnarray}
where $g(\epsilon)$ is the density of states (\ref{12r}), the function
$ R^{(\nu_1,\nu_2)}(\epsilon,\epsilon_1,\omega) $ is given by 

\begin{equation}
 R^{(\nu_1,\nu_2)}(\epsilon,\epsilon_1,\omega)=
2\sum_{l=1}^{+ \infty} I_l(z) 
\sinh \left( \frac{\beta \bar{\omega}_0 l}{2} \right)
\left[ 
J_l^{(\nu_1,\nu_2)}(\epsilon,\epsilon_1,\omega)+
H_l^{(\nu_1,\nu_2)}(\epsilon,\epsilon_1,\omega)
\right],
\label{135ar}
\end{equation}
$C^{(\nu_1,\nu_2)}(\epsilon,\epsilon_1,x)$ is   

\begin{equation}
 C^{(\nu_1,\nu_2)}(\epsilon,\epsilon_1,x)=
\frac{1}{2}
\frac{ [\Gamma(\epsilon)+\Gamma(\epsilon_1)] }
{ [\Gamma(\epsilon)+\Gamma(\epsilon_1)]^2/4+
(\xi^{(\nu_1)}-\xi_1^{(\nu_2)}+x)^2 }
\label{131r}
\end{equation} 
and 

\begin{equation}
\xi_{i}^{(\nu_{j})}=\nu_{j} \sqrt{\epsilon^2_{i}+E^2}.
\end{equation}
We notice that the functions 
$J_l^{(\nu_1,\nu_2)}( \epsilon,\epsilon_1,\omega )$ 

\begin{eqnarray}
J_l^{(\nu_1,\nu_2)}( \epsilon,\epsilon_1,\omega ) &=&
C^{(\nu_1,\nu_2)}( \epsilon,\epsilon_1,\omega+l\bar{\omega}_0 ) 
[n_F(\xi^{(\nu_2)}_1-l\bar{\omega}_0-\omega)-
n_F(\xi^{(\nu_2)}_1-l\bar{\omega}_0)]
\times
\nonumber \\
&&
\times
\left[ N_0(l \bar{\omega}_0)+n_F(\xi^{(\nu_2)}_1) \right]
\label{136ar}
\end{eqnarray}
and $H_l^{(\nu_1,\nu_2)}( \epsilon,\epsilon_1,\omega )$ 
 
\begin{eqnarray}
H_l^{(\nu_1,\nu_2)}( \epsilon,\epsilon_1,\omega ) &=&
C^{(\nu_1,\nu_2)}( \epsilon,\epsilon_1,\omega-l\bar{\omega}_0 )
[n_F(\xi^{(\nu_2)}_1+l\bar{\omega}_0-\omega)-
n_F(\xi^{(\nu_2)}_1+l\bar{\omega}_0)]
\times
\nonumber \\
&&
\times
\left[ N_0(l \bar{\omega}_0)+1-n_F(\xi^{(\nu_2)}_1) \right]
\label{137ar}
\end{eqnarray} 
describe phonon absorption and emission processes, respectively. 

In the limit of high temperatures and for small polaron excitations, the 
phononic incoherent absorption is prevalent and in this limit an analytic 
expansion can be performed. \cite{perroni} 

The internal consistency of our approach is verified by using the
sum rule
\begin{equation}
\int_0^{\infty} d\omega \Re \sigma_{\alpha,\alpha}(\omega)=
- \frac{\pi}{2} e^2 <\hat{T}_{\alpha,\alpha}>   
\label{560r}
\end{equation}
where $\langle \hat{T}_{\alpha,\alpha} \rangle $ is the mean value of 
component of the kinetic energy equal to one third of eq.(\ref{562ra}).
Employing the calculated spectra, we have checked tha the two sides of 
eq.(\ref{560r}) differ in a few per cent.  

In Fig.6 we report the conductivity spectra for $T=0.023 W$ at 
$el-ph$ couplings.
With rising the $el-ph$ coupling, a tranfer of spectral weight from
low to high energies takes place.
The Drude term makes smaller and, in the weak to intermediate 
$CO$ regime, the optical response is characterized by two components: the 
remnant Drude-like term and the incipient  absorption band centered
around the gap energy.
For stronger $el-ph$ couplings, the band is peaked around the gap but 
it is accompained by a subgap tail due to the lattice fluctuations.
\cite{ross,degiorgi1,wilkins}
Thus, by increasing the parameter $\lambda$ and by crossing the phase 
transition, we do not have any sharp changes in the optical spectra but a 
rather continuous evolution.
We point out that in the mean-field approach the infrared absorption occurs
only at energies above the gap and it is characterized by the inverse 
square-root singularity.
\cite{anderson}

In the inset of Fig.6 the gradual disappearance of the Drude term and the rise
of the interband absorption can be followed in a narrow range of energies.
At higher temperatures a similar evolution of the bands of the
infrared absorption is found by increasing the $el-ph$
strength. Actually, in both the ranges of temperature, the response in
the limit $\omega \rightarrow 0$ is completely suppressed only for
strong $el-ph$ couplings where a well-defined optical gap opens.  
Furthermore in the $CO$ phase there is a lowering of the height of the
interband terms that spread over a wide range of energies.

It is interesting to analyze the conductivity spectra at a fixed $el-ph$ 
coupling for different temperatures.
We take into account two different regimes: $\lambda=0.4$ ($A \rightarrow 
 CO$ phase) and $\lambda=0.8$ ($B \rightarrow CO$ phase).
In the first case (Fig.7), by starting from the normal phase and by lowering 
the temperature, there is a redistribution of the spectral weight among the
two components accompanied by a progressive narrowing of the Drude term.
The maximum of the second band slightly shifts towards higher frequencies
and the optical gap fully develops at low temperatures well within the $CO$
phase.
We stress that these results of the optical absorption are consistent with
experimental measurements in dichalcogenides.
\cite{vescoli}

In the other regime ($\lambda=0.8$) very different features of the optical 
response are obtained (Fig.8). Here the evolution from the $SP$ to $CO$ 
absorption band is characterized by near peak energy. In the $CO$ phase, 
below $T_{CO} $, there is a residue of optical response in the limit
 $\omega \rightarrow 0$ and the optical gap starts to appear for lower 
temperatures in corrispondence of energies that can decrease with lowering $T$.
These results can be closely connected with the conductivity spectra measured
in nickelates \cite{katsu,jung} that reveal the relevance of the lattice 
degrees in stabilizing the $CO$ phase.

\section{Summary and conclusions} 

We have discussed the phase diagram, the spectral and optical properties of 
the half-filled spinless Holstein model in three dimensions as a function of 
the temperature and the $el-ph$ coupling. 
The lattice fluctuation effects play a crucial role in determining 
the density of states and the optical response of the system.
  
The phase diagram is consistent with previous $DMFT$ results pointing out a 
strong dependence of the transition temperature on the adiabaticity 
particularly in the weak to intermediate coupling regime. The density
of states captures the features of the intermediate coupling regime in
agreement with $DMFT$ studies.

Concerning the optical absorption, we have observed that, with decreasing 
$T$ or increasing the $el-ph$ coupling, the ordered phase affects the 
conductivity spectra inducing a transfer of spectral weight from low to high 
energies with the opening of an optical gap.
The inclusion of effects of quantum lattice fluctuations is able to smooth 
the inverse square-root singularity of the mean-field approach and 
induces a subgap tail absorption.
In the weak to intermediate coupling regime, a two-peak structure is 
obtained within the $CO$ phase at low frequency: 
the remnant Drude-like term and the incipient absorption band centered around 
the gap energy.
The results obtained in the intermediate and in the strong coupling regime 
are consistent with experimental conductivity spectra.
\cite{katsu,jung,vescoli,degiorgi1}

In this paper lattice fluctuation effects beyond the variational $ZFA$
approach are included by means of self-energy insertions. At the
lowest order the main contribution to the self-energy is independent
on the momentum, however at the same order
other terms could be easily included in order to give a non local
quantity. Furthermore, the scattering rate can be calculated not only
perturbatively but also employing a self-consistent procedure. 
\cite{schna,loos1,loos2,perroni}
If also the real part of the self-energy is introduced by using the
Kramers-Kronig relation, the procedure of calculation could become
more complete. The agreement of the results 
obtained in this paper with other approaches seems to suggest that a
second-order perturbation theory on the good $ZFA$ solution can
capture the physics of the model in the different regimes of the
normal and ordered state.   
   
Finally we note that our approach is valid in the infrared range of 
frequencies where the interband absorption takes place. Thus it is not able 
to reveal the structures attribuited to collective excitation modes arising 
from the $CDW$ condensate. \cite{gruner,gruner1,degiorgi,kida}

\section*{Figure captions}
\begin{description}

\item  {F1} 
The phase diagram (solid line) corresponding to $W=6\omega_0$. $CO$ 
characterizes the charge ordering phase, the $A$ phase represents the 
Fermi-liquid-like normal state, the $B$ phase the disordered
small polaron normal state (the crossover between the $A$ and $B$ phase is
marked by the dashed line). In the inset, the ratio of $T=0$ gap 
$\Delta$ to the ordering temperature $T_{CO}$ as a function of $\lambda$.

\item  {F2} 
(a) The transition temperature as a function of $\lambda$ for two different 
values of the adiabaticity ratio: $W=6\omega_0$ (solid line) and $W=3\omega_0$
(dashed line).
(b) The variation $\Delta T_{CO}=T_{CO}^{*}-T_{CO}$ of the transition 
temperatures due to different masses of the oscillator as a function of 
$\lambda$.

\item  {F3} 
(a) The renormalized density of states at $W=6\omega_0$ and $T=0.023 W$  
as a function of the energy (in units of $\omega_0$) for different $el-ph$ 
couplings: $\lambda=0.5$ (solid line), $\lambda=0.35$ (dashed line) and 
$\lambda=0.15$ (dotted line). 

(b)The renormalized density of states at $W=6\omega_0$ and $T=0.14 W$  
as a function of the energy (in units of $\omega_0$) for different $el-ph$ 
couplings: $\lambda=0.65$ (solid line), $\lambda=0.6$ (dashed line), 
$\lambda=0.56$ (dotted line) and $\lambda=0.4$ (dash-dotted line). 

\item  {F4} 
(a) The scattering rate at $W=6\omega_0$ and $T=0.023 W$  
as a function of the energy (in units of $\omega_0$) for different $el-ph$ 
couplings: $\lambda=0.5$ (solid line) and $\lambda=0.15$ (dashed line).

(b) The scattering rate at $W=6\omega_0$ and $\lambda=0.4 W$  
as a function of the energy (in units of $\omega_0$) for different temperatures
: $T=0.01 W$ (solid line), $T=0.05 W$ (dashed line), and $T=0.14 W$ 
(dotted line). 

The arrows indicate the gap energy.

\item  {F5} 
(a) The new renormalized density of states at $W=6\omega_0$ and $\lambda=0.2$  
as a function of the energy (in units of $\omega_0$) for different 
temperatures. 

(b)The new renormalized density of states at $W=6\omega_0$ and $T=0.2 W$  
as a function of the energy (in units of $\omega_0$) for different $el-ph$ 
couplings.

\item  {F6} 
The conductivity up to 6 $\omega_0$ at different $el-ph$ couplings.
In the inset the conductivity up to 2 $\omega_0$ at different $el-ph$ 
couplings.
The conductivities are expressed in units of $\frac{e^2 \rho}{m t}$, with
$m= \frac {1}{2t}$.

\item  {F7}
The conductivity (in units of $\frac{e^2 \rho}{m t}$, with
$m= \frac {1}{2t}$) up to 2 $\omega_0$ at different $el-ph$ couplings.

\item  {F8} 
The conductivity (in units of $\frac{e^2 \rho}{m t}$, with
$m= \frac {1}{2t}$) up to 16 $\omega_0$ at different $el-ph$ couplings.

\end{description}

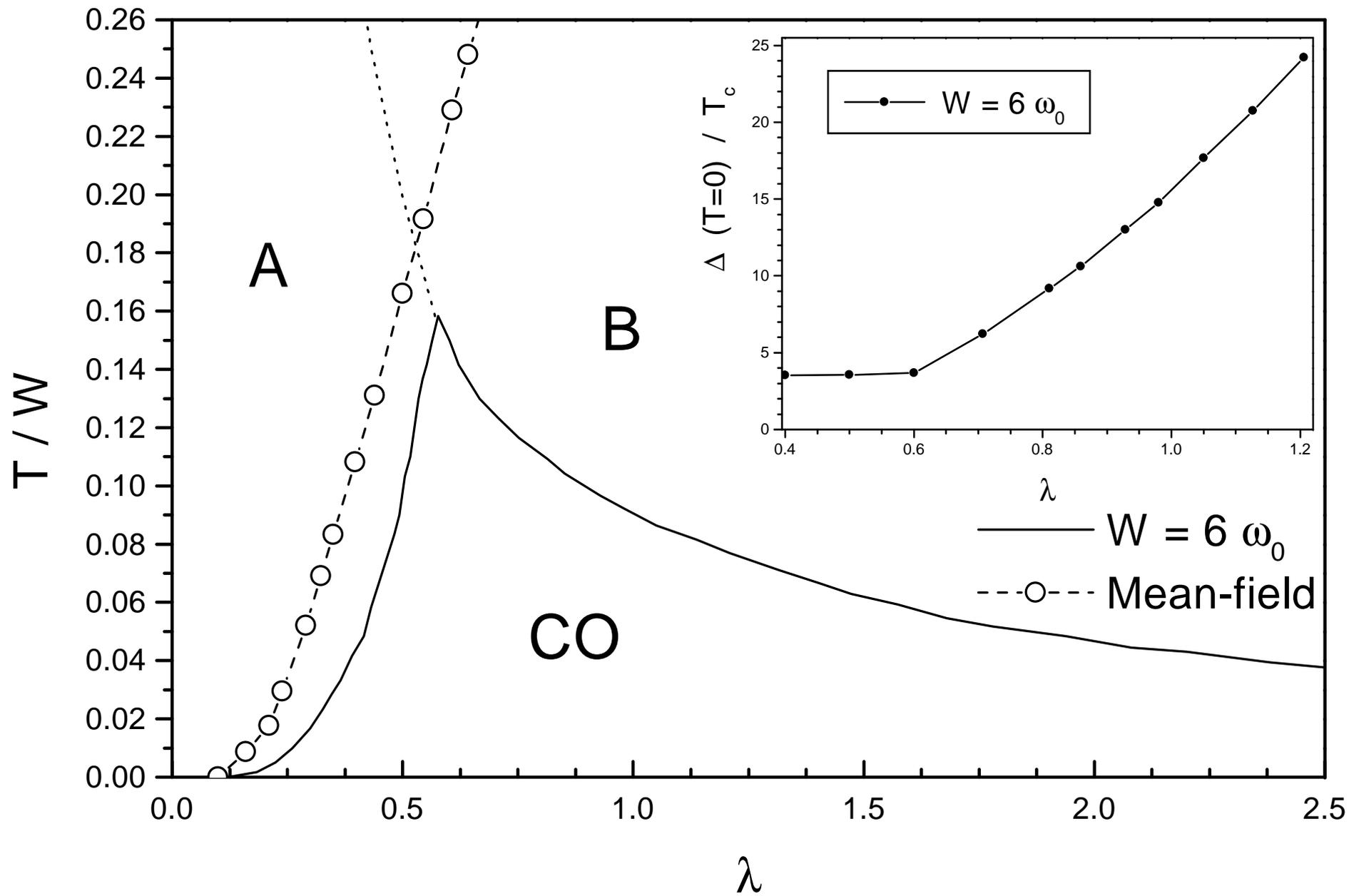
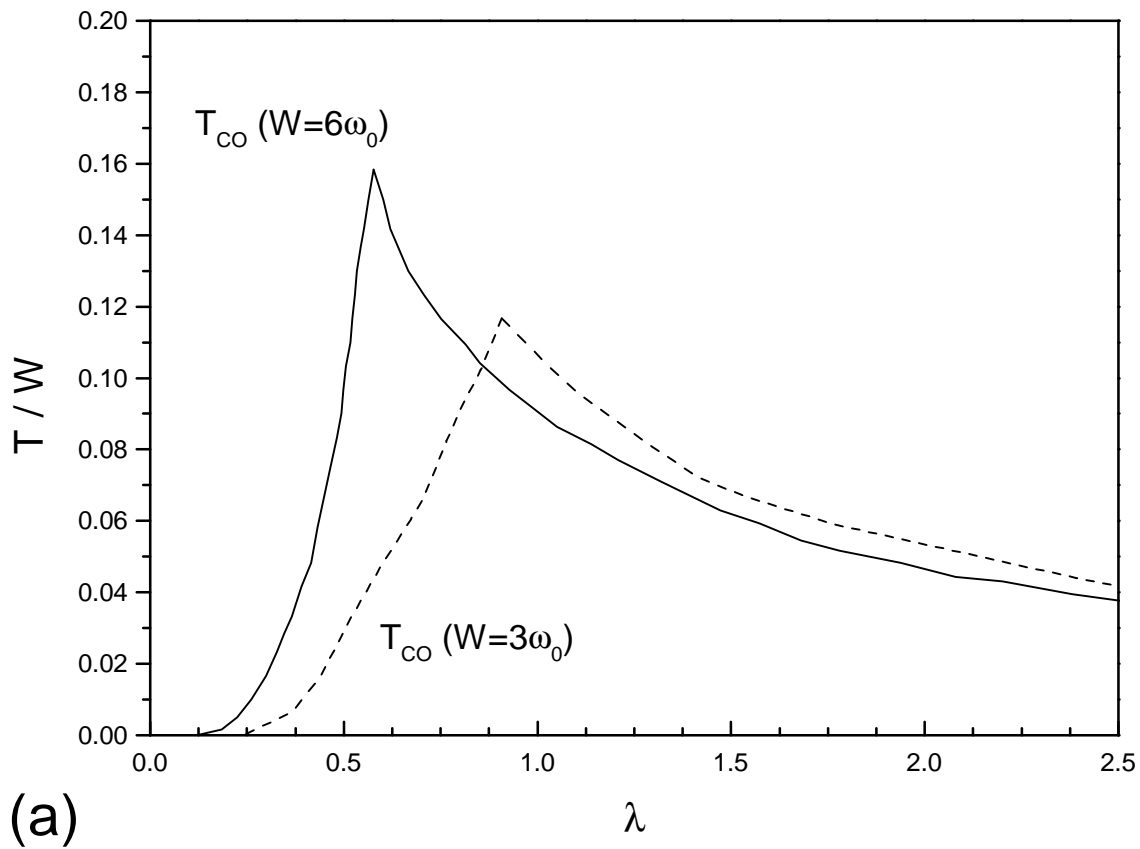
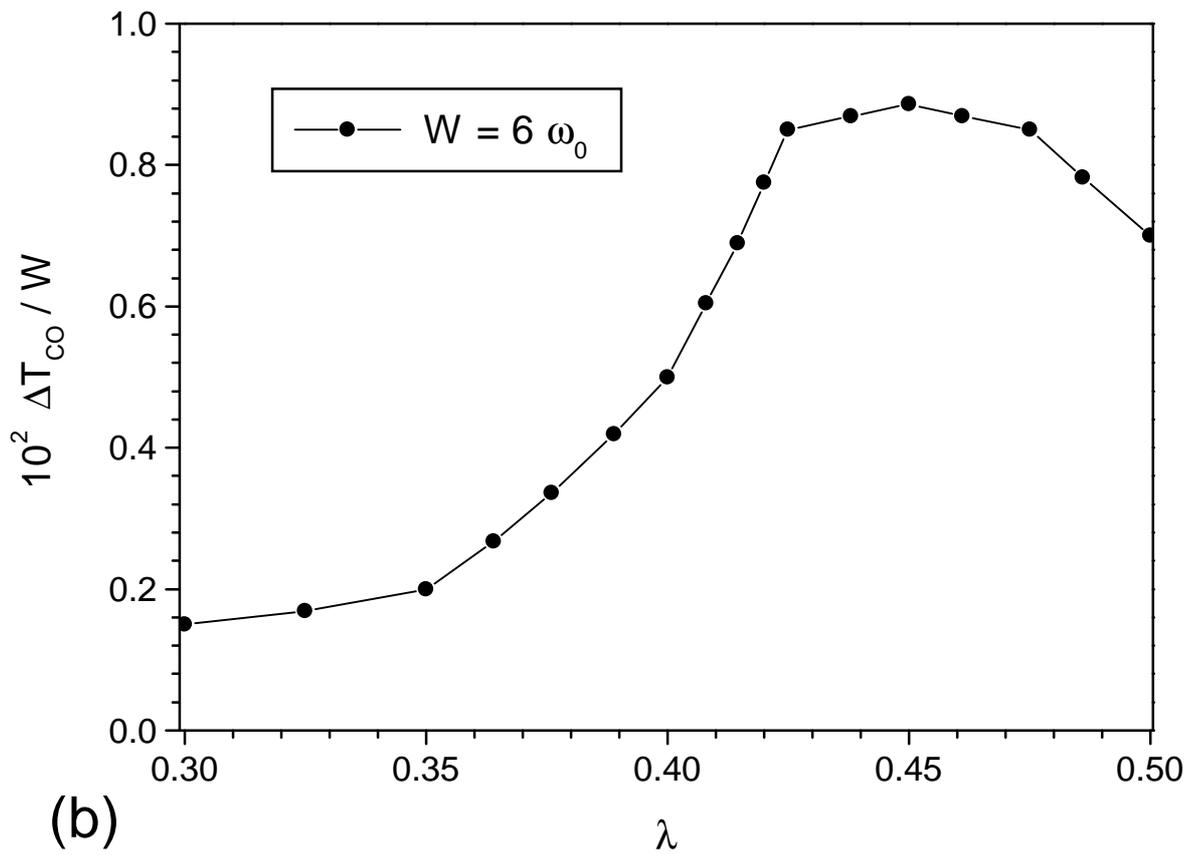
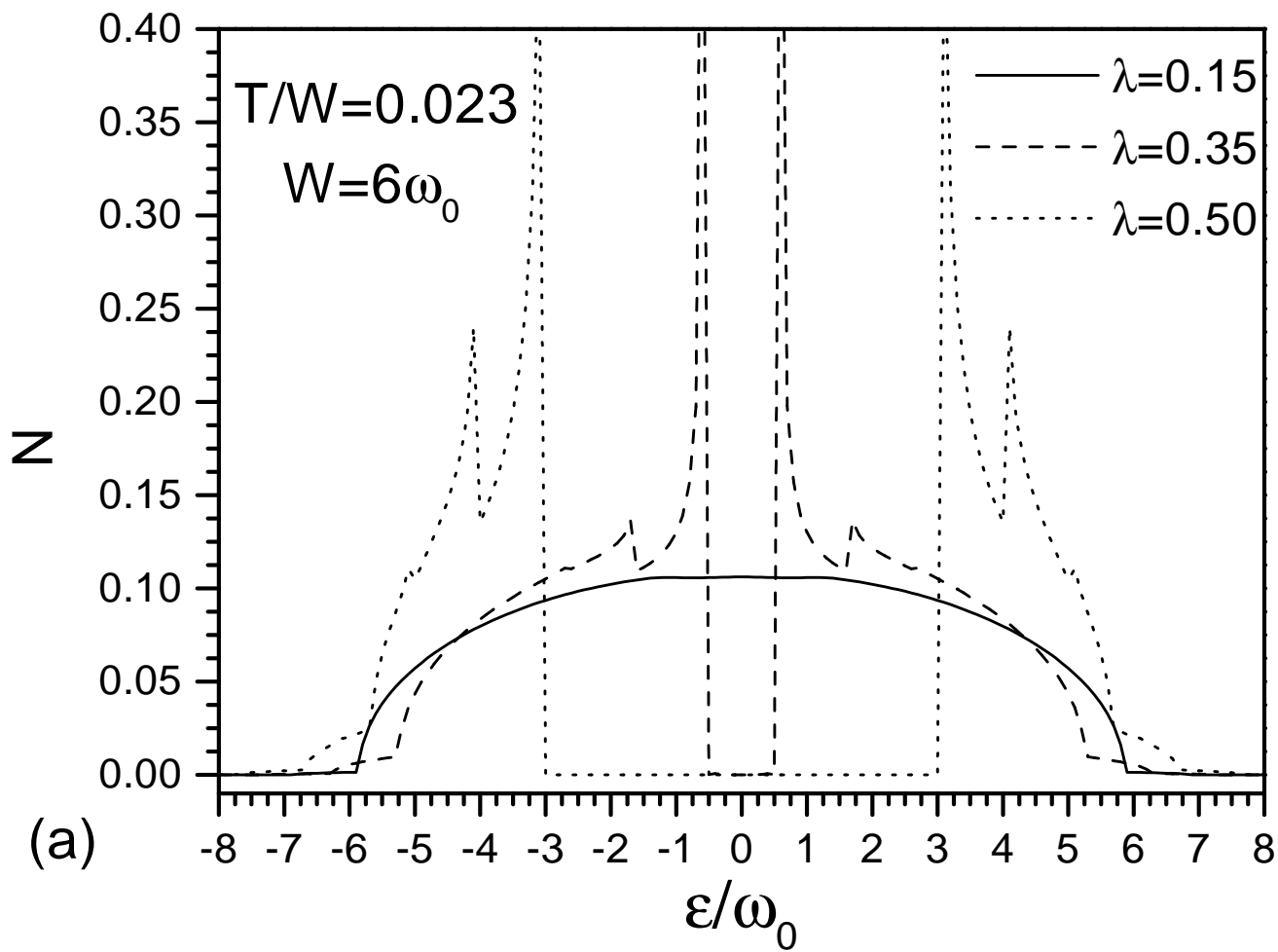
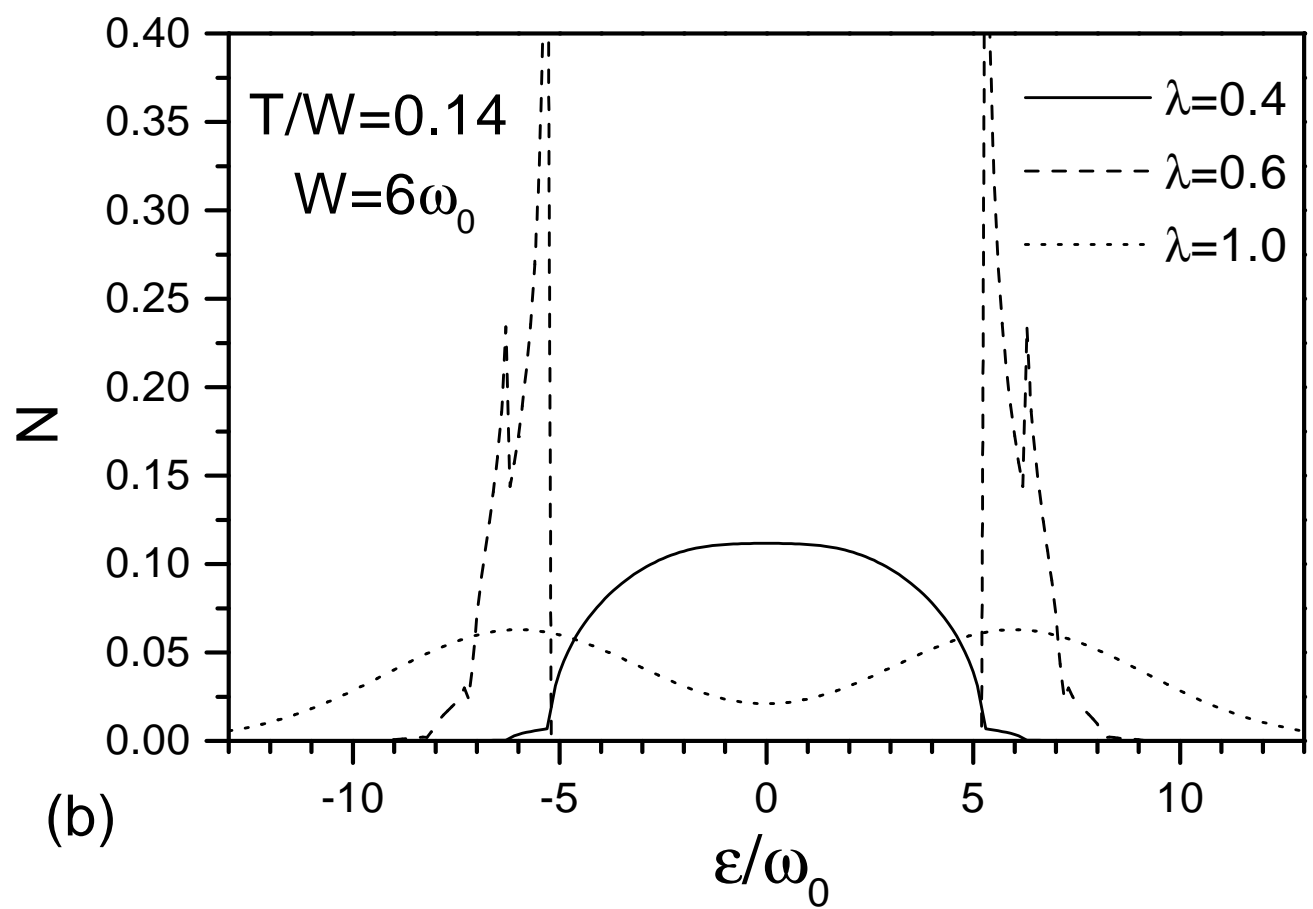
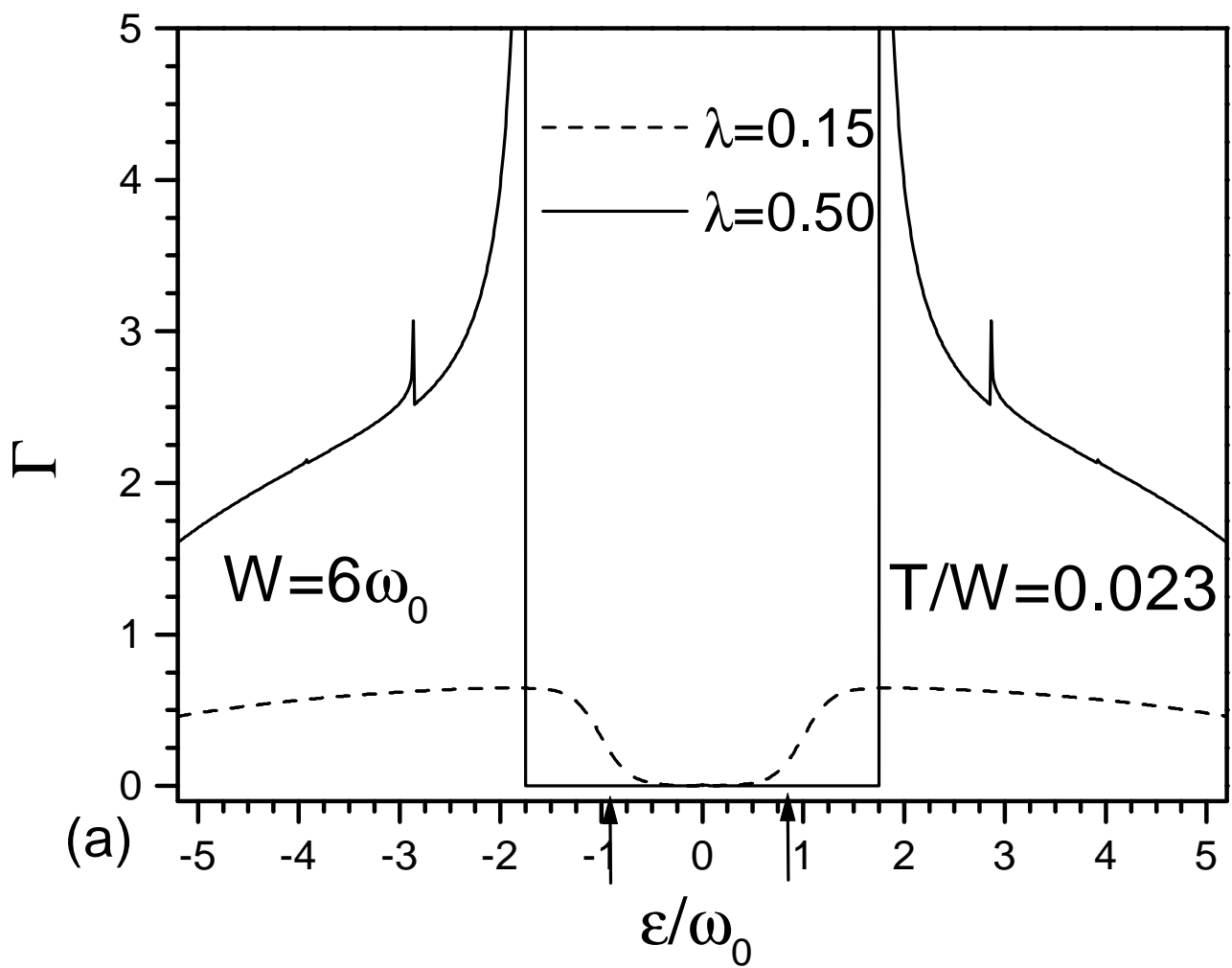
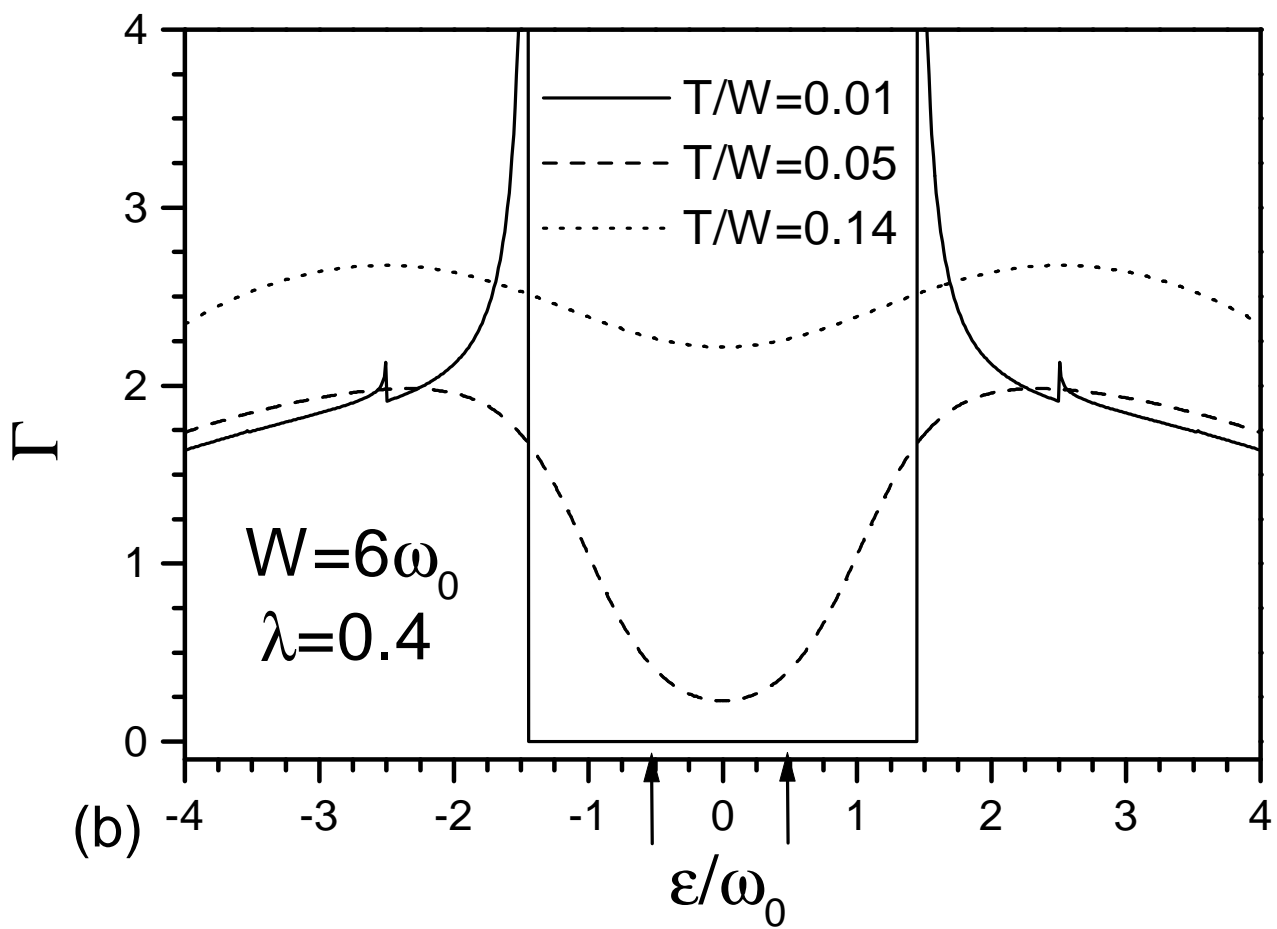
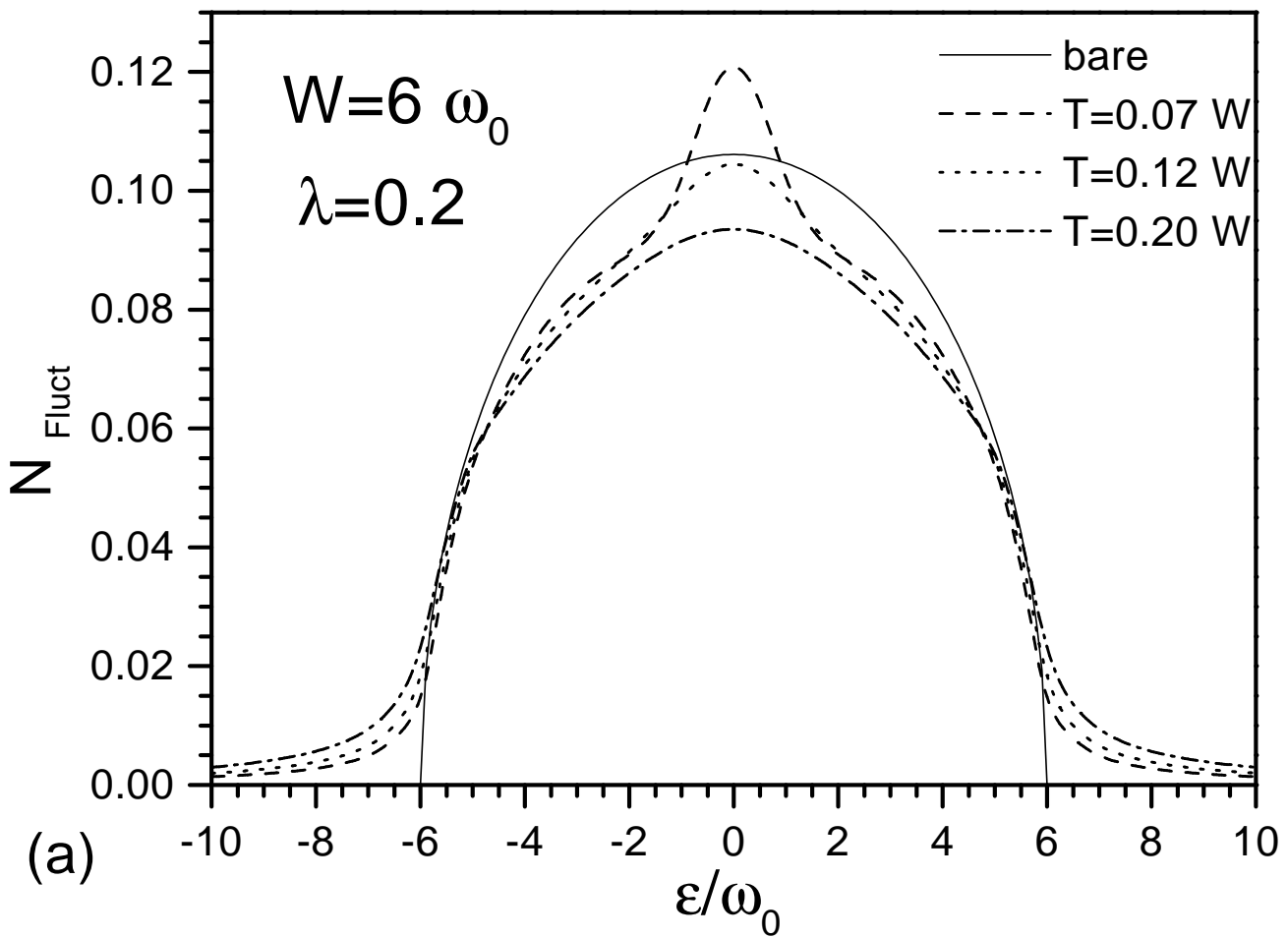
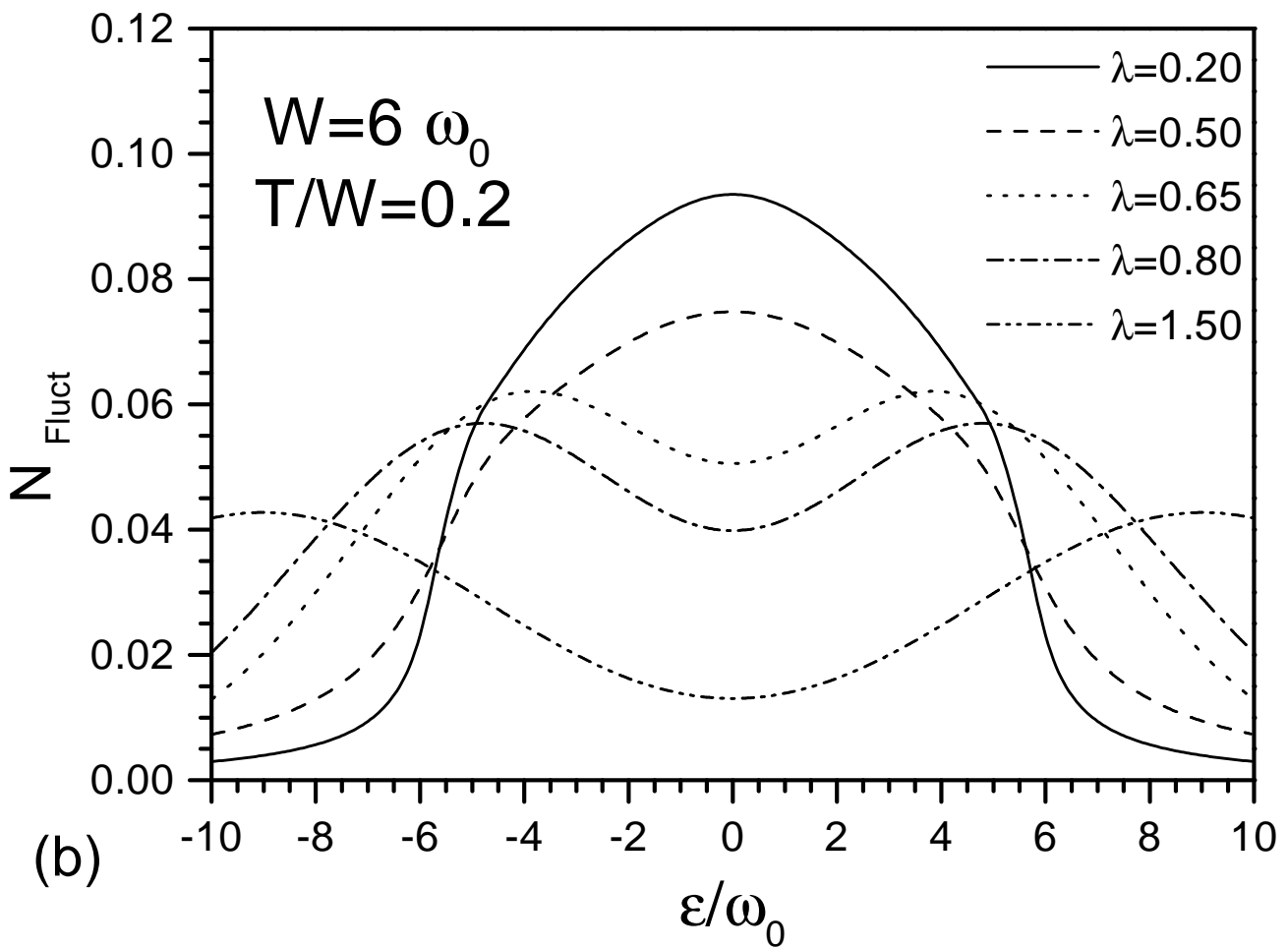
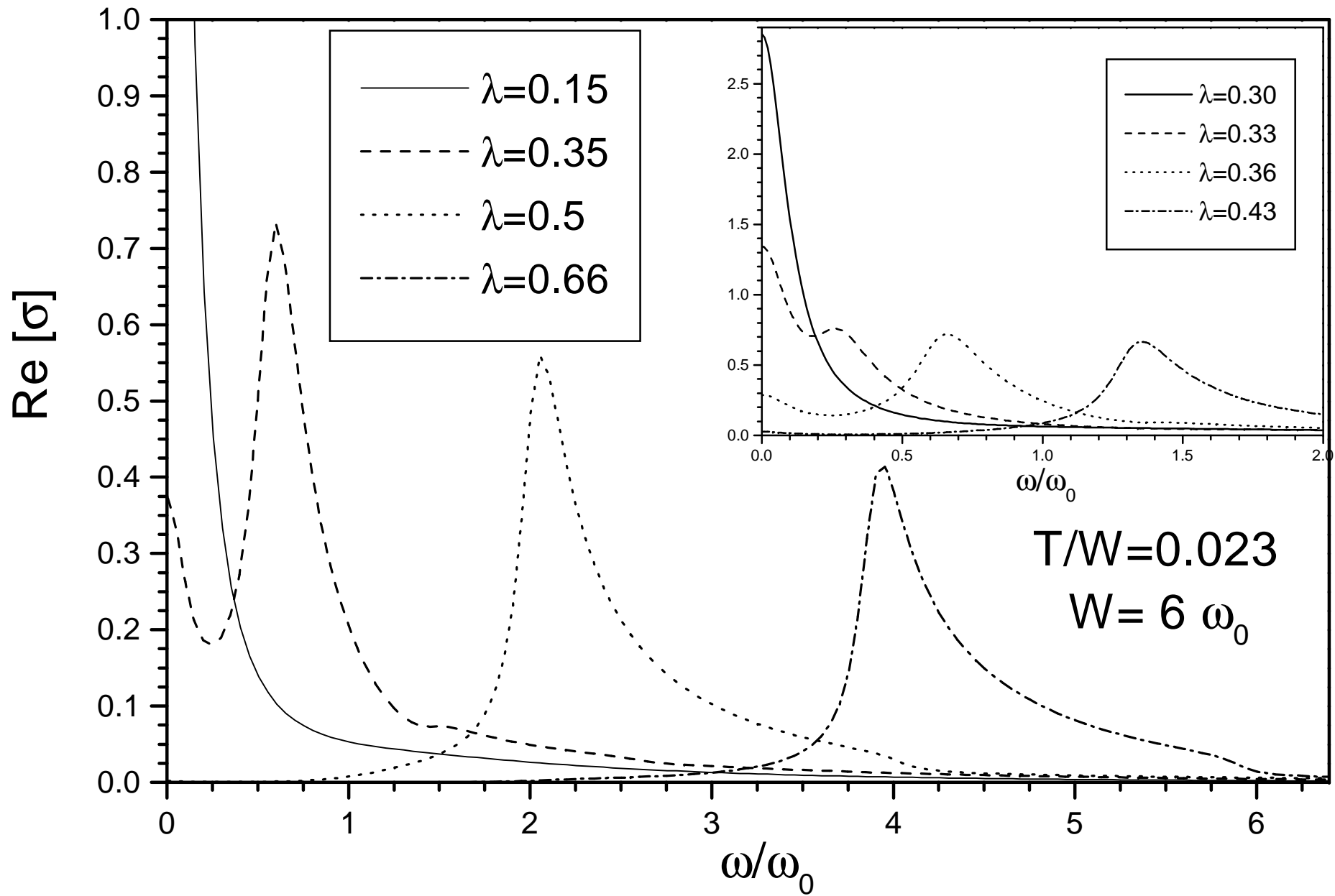
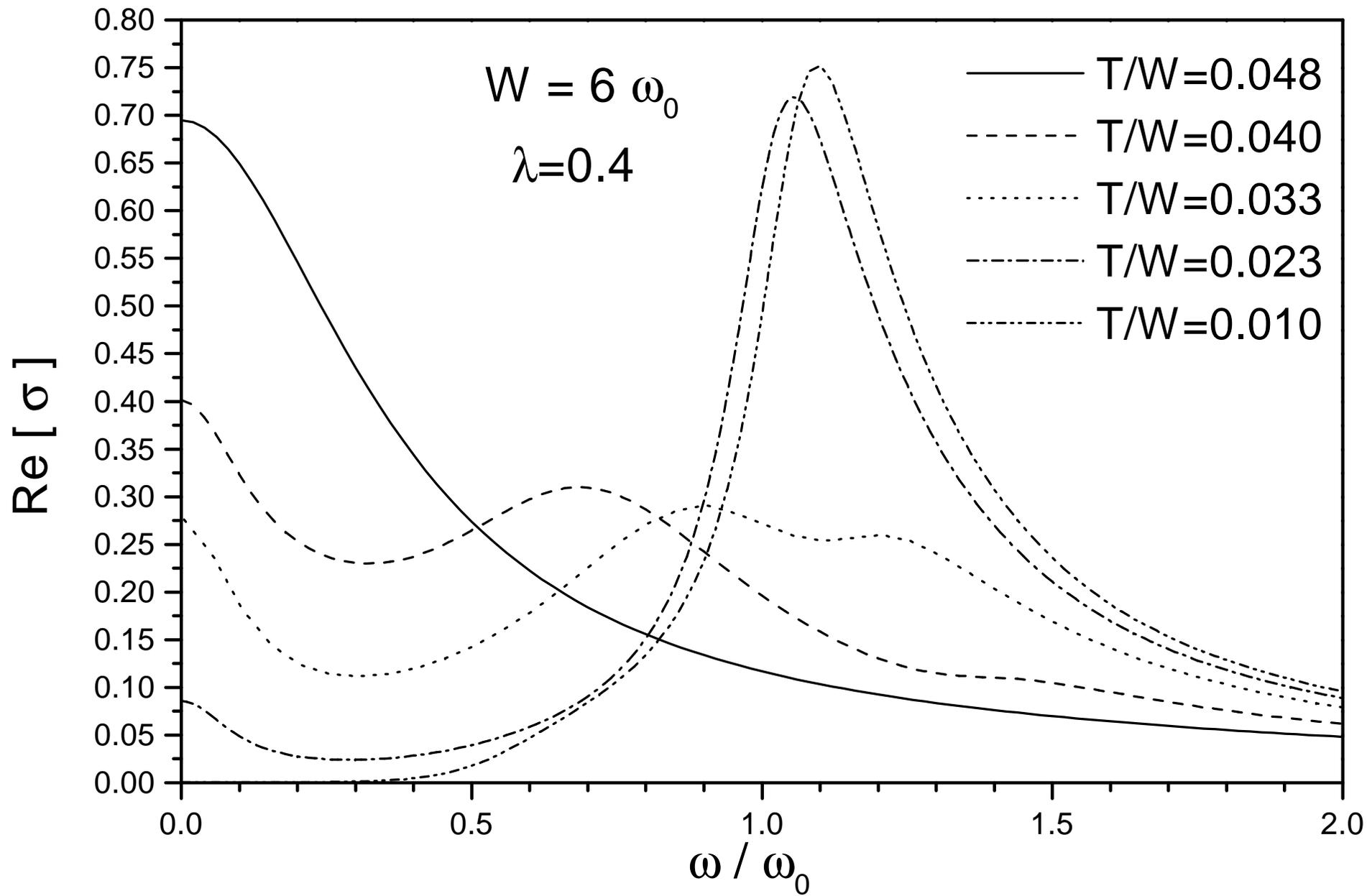
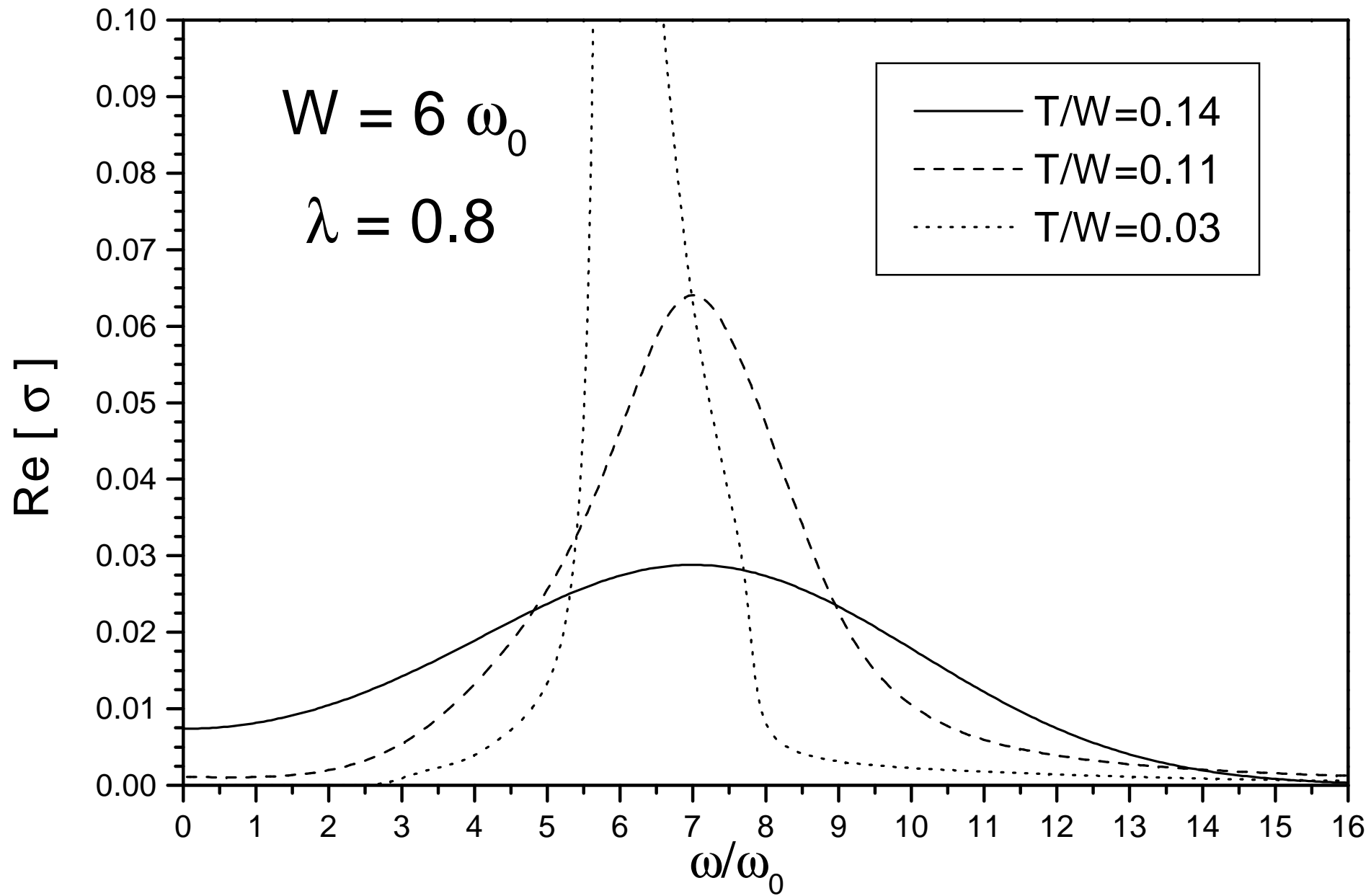
\end{document}